\documentclass[aps,prd,reprint]{revtex4-2}
\usepackage{graphicx}
\usepackage{dcolumn}
\usepackage{bm}
\usepackage{amsmath}
\usepackage{amssymb}
\usepackage{textcomp}
\usepackage{graphicx}
\usepackage{caption}
\usepackage{subcaption}
\usepackage{array}
\usepackage{mathtools}
\usepackage{centernot}
\usepackage{slashed}
\usepackage{hyperref}
\usepackage{xcolor}
\usepackage{tikz-feynman}
\usepackage{mhchem}
\usepackage{booktabs}
\usepackage{multirow}
\usepackage{rotating}
\usepackage{tabularx}
\usepackage{siunitx}
\usepackage{placeins}
\usepackage{float}

\hypersetup{
	colorlinks=true,       
	linkcolor=blue,          
	citecolor=magenta,        
	filecolor=cyan,      
	urlcolor=magenta           
}

\begin{document}
	\title{Signatures of Type-I Seesaw in Neutrino Oscillation Phenomenology}
\author{Suka Sriyansu Pattanaik}
\email{421ph5072@nitrkl.ac.in}
\author{Sasmita Mishra}
\email{mishras@nitrkl.ac.in}
\affiliation{Department of Physics and Astronomy, National 
Institute of Technology Rourkela, Sundargarh, Odisha, India, 769008}
	
\begin{abstract}
We investigate the low-energy phenomenology of the Type-I seesaw mechanism within a 3+3 framework containing three active and three sterile neutrinos. Using the exact seesaw relation as a bridge between the high-scale sterile-sector parameters and the standard oscillation observables, we perform a comprehensive Monte Carlo scan of the 21-dimensional sterile parameter space, retaining only those configurations consistent with current neutrino oscillation data within $3\sigma$. For the viable parameter points, we simulate the modified neutrino oscillation probabilities and event rates at the long-baseline experiments DUNE and NO$\nu$A, and the medium-baseline reactor experiment JUNO, quantifying their sensitivity to sterile neutrino effects across the eV--GeV mass range. We find that eV-scale sterile neutrinos produce pronounced spectral distortions, while heavier states decouple progressively from oscillation experiments. In parallel, we confront the seesaw predictions with complementary probes: cosmological bounds on $\sum m_i$, the kinematic mass $m_\beta$ from beta decay, the effective Majorana mass $|m_{\beta\beta}|$ from neutrinoless double beta decay ($0\nu\beta\beta$), and the charged-lepton-flavor-violating branching ratio $\text{BR}(\mu \to e\gamma)$. The combination of all constraints significantly narrows the allowed parameter space: the predicted sum of neutrino masses clusters at $\sum m_i \sim 0.05$--$0.07$~eV, within reach of next-generation cosmological surveys, and eV-scale sterile neutrinos are found to be under significant tension from the current MEG bound on $\mu \to e\gamma$.
\end{abstract}

\maketitle


\section{Introduction}
\label{sec:intro}
The discovery of neutrino oscillations~\cite{Super-Kamiokande:1998kpq,SNO:2002tuh} provides irrefutable evidence that neutrinos are massive particles, a direct contradiction to the predictions of the Standard Model (SM) where they are considered massless. This observation stands as one of the most significant empirical proofs of physics beyond the Standard Model (BSM) and raises fundamental questions regarding the mass generation mechanism, the absolute mass scale, and the intrinsic properties of neutrinos~\cite{Gonzalez-Garcia:2002bkq,Maltoni:2003da}. To accommodate neutrino mass, the SM Lagrangian must be extended. 
Among the most compelling and economical extensions is the Type-I seesaw mechanism~\cite{Minkowski:1977sc,Yanagida:1979as,Gell-Mann:1979vob,Mohapatra:1979ia,Schechter:1980gr}, which introduces right-handed neutrinos ($N_R$) that are singlets under the SM gauge group and therefore do not participate in gauge interactions, hence the designation \textit{sterile} neutrinos. 
These new states possess a Majorana mass term, and the resulting seesaw relation naturally explains the smallness of the active neutrino masses. However, this typically places the new physics at an energy scale far too high to be probed directly, making it difficult to test this elegant hypothesis. 

Beyond their role in neutrino mass generation, sterile neutrinos have far-reaching implications in cosmology and particle physics: they can provide viable dark matter candidates~\cite{Boyarsky:2009ix,Asaka:2005an}, generate the observed baryon asymmetry of the Universe via leptogenesis~\cite{Fukugita:1986hr,Akhmedov:1998qx}, and contribute to lepton-number-violating processes such as neutrinoless double beta decay~\cite{Bolton:2019pcu,Dolinski:2019nrj}. A comprehensive review of the phenomenology of right-handed neutrinos can be found in Refs.~\cite{Drewes:2013gca,Abdullahi:2022jlv}. The search for sterile neutrinos has been intensely driven by persistent experimental anomalies, most notably from the LSND~\cite{LSND:2001aii} and MiniBooNE~\cite{MiniBooNE:2018esg} experiments, although recent results from MicroBooNE~\cite{MicroBooNE:2021tya} have not confirmed the MiniBooNE low-energy excess under a simple $\nu_e$ interpretation. Nevertheless, these results hint at the existence of one or more light sterile neutrinos at the eV-scale, which would participate in oscillations with the active flavors. This has led to phenomenological (3+N) models that can be embedded within more complete low-scale seesaw frameworks~\cite{deGouvea:2022kma,Fernandez-Martinez:2016lgt}. Such models are particularly attractive as they offer a rich, testable phenomenology at accessible energy scales while remaining theoretically grounded in the seesaw framework.

%
The role of sterile neutrinos, however, extends well beyond merely suppressing the active neutrino mass scale through the seesaw relation. Their effects on low-energy observables arise through multiple, distinct mechanisms whose interplay shapes \textit{all} measurable properties of the active neutrinos. As additional mass eigenstates, sterile neutrinos introduce new propagating degrees of freedom that \textit{directly} participate in physical processes: at the eV-scale, they contribute additional oscillation frequencies that alter appearance and disappearance probabilities at short- and long-baseline experiments~\cite{deGouvea:2022kma}, while in charged lepton flavor violating (cLFV) processes such as $\mu \to e\gamma$, the heavy sterile states enter as virtual mediators in quantum loop diagrams, generating rates that depend directly on the sterile masses and their couplings to the active flavors~\cite{Ilakovac:1994kj,Abada:2015oba}. Independently, when the sterile states are sufficiently heavy that their oscillations average out at a given baseline, they leave an \textit{indirect} imprint through a non-unitary effective $3\times 3$ PMNS matrix~\cite{Antusch:2006vwa,Escrihuela:2015wra,Parke:2015goa}. These non-unitarity corrections modify the effective values of the mixing angles, mass-squared differences, and the CP-violating phase~\cite{Blennow:2016jkn,Fernandez-Martinez:2016lgt,Blennow:2023mqx}, potentially mimicking or obscuring the determination of $\delta_{CP}$~\cite{Klop:2014ima,Escrihuela:2015wra} and shifting the apparent values of $\theta_{23}$ and $\theta_{13}$ at long-baseline facilities~\cite{deGouvea:2022kma}. In practice, the observable signatures in any given experiment reflect an intertwined combination of both these direct and indirect effects, whose relative importance depends on the sterile mass scale and mixing. On the cosmological and phenomenological fronts, the absolute neutrino mass scale---probed through the sum of neutrino masses $\sum m_i$~\cite{Planck:2018vyg}, the effective mass in neutrinoless double beta decay $|m_{ee}|$~\cite{Das:2023aic}, and the kinematic endpoint in beta decay $m_\beta$~\cite{KATRIN:2024cdt}---is itself a prediction of the seesaw parameters and is shaped by the interplay of sterile states with the active sector. The question, therefore, is not only \textit{whether} sterile neutrinos exist but how the combined effect of direct sterile neutrino contributions and indirect non-unitarity corrections shapes the active neutrino phenomenology that current and near-future experiments are designed to measure.

In this work, we address this question within a 3+3 Type-I seesaw framework. Using the exact seesaw relation~\cite{Xing:2024gmy} as the bridge between the high-scale sterile parameters and the low-energy active sector, we perform a comprehensive scan of the sterile-sector parameter space to map the full range of effects that seesaw-induced sterile neutrinos can produce on the standard oscillation observables. We explicitly simulate the modified neutrino oscillation probabilities at the Long-Baseline (LBL) experiments DUNE~\cite{DUNE:2025sjq} and NO$\nu$A~\cite{NOvA:2007rmc}, and the Medium-Baseline (MBL) reactor experiment JUNO~\cite{JUNO:2021vlw}, assessing how the sensitivity of each experiment is affected across different sterile mass scales. Alongside the oscillation analysis, we derive predictions for the absolute neutrino mass scale and confront them with constraints from cosmology, $0\nu\beta\beta$ decay searches, kinematic measurements, and charged lepton flavor violation, thereby providing a unified picture of the observational consequences of the seesaw mechanism at current and upcoming experiments.

The remainder of this paper is organized as follows. In Sec.~(\ref{sec:seesaw_mechanism}), we introduce the Type-I seesaw mechanism, the parametrization of the $6\times 6$ mixing matrix, and derive the exact seesaw relation that connects the sterile-sector parameters to the active neutrino observables. In Sec.~(\ref{sec:seesaw_oscillations}), we discuss the modifications to neutrino oscillation probabilities arising from the seesaw-induced active-sterile mixing and present the simulation results for DUNE, NO$\nu$A, and JUNO. In Sec.~(\ref{subsec:complementarity_probes}), we examine the complementarity between oscillation experiments and other probes of neutrino mass, including cosmological constraints, $0\nu\beta\beta$ decay, kinematic mass measurements, and charged lepton flavor violation. In Sec.~(\ref{sec:low_scale_justification}) we comment on some model-building aspects of low-scale seesaw models. Finally, we present our conclusions in Sec.~(\ref{sec:conclusion}).

\section{Type-I Seesaw Mechanism}
\label{sec:seesaw_mechanism}
To realize the Type-I seesaw, the SM is extended with three right-handed sterile neutrinos $N_{Ri}, i =1,2,3$. 
The relevant Lagrangian is of the form
\begin{equation}
    -\mathcal{L} \supset i\overline{N_{Ri}}\gamma^\mu \partial_\mu N_{Ri} - (Y_{\alpha i} \overline{L_\alpha} \tilde{\Phi} N_{R i} + \frac{1}{2}M_{Ri}\overline{N^c_{Ri}}N_{Ri}+h.c.),
\end{equation}
where $\Phi$ is the SM Higgs doublet, $\left( \phi^+ ~~\phi^0 \right)^T$ with $\tilde{\Phi} = 
i\sigma_2 \Phi^*$, and  $L_\alpha$ represents the lepton doublet 
$\left( \nu_{L\alpha} ~ e_{L\alpha} \right)^T$, where $\alpha = e, \mu, \tau$. $Y_{\alpha i}$ is the Yukawa coupling matrix of the corresponding $L_\alpha$ with $N_R$. The Majorana mass matrix $M_R$ is taken to be positive and real. After electroweak symmetry breaking, the Higgs field acquires a non-zero vacuum expectation value $\langle \Phi \rangle = (0 \;\; v/\sqrt{2})^T$, where $v \simeq 246$~GeV.
The Yukawa interaction then generates a Dirac mass term coupling the active
and sterile neutrinos:
\begin{equation}
    m_D = \frac{Y_\nu \, v}{\sqrt{2}},
    \label{eq:dirac_mass}
\end{equation}
where $Y_\nu$ is the $3\times 3$ neutrino Yukawa coupling matrix.
The resulting neutrino mass Lagrangian takes the form
\begin{equation}\label{eq:mass_lagrangian}
    \mathcal{L}_{mass} \supset -\overline{\nu_L} m_D N_R
    + \frac{1}{2}\overline{N^c_{Ri}} M_{Ri} N_{Ri} + \text{h.c.}
\end{equation}
Writing the mass terms in the combined basis
$\Psi^T = (\nu_L \;\; N_R^c)$, the full $6\times 6$ neutrino mass matrix takes the block form
\begin{equation}\label{eq:mass_Matrix}
    \mathcal{M} = \begin{pmatrix} 0 & m_D \\ m_D^T & M_R \end{pmatrix}
\end{equation}
The null matrix in the upper-left block reflects the absence of a bare Majorana mass term for the left-handed neutrinos in the SM, while the off-diagonal Dirac mass $m_D$ couples the active and sterile sectors. The Majorana mass $M_R$ in the lower-right block is the bare mass of the sterile neutrinos. Being a Majorana mass term, it violates total lepton number by two units ($\Delta L = 2$) and is responsible for the Majorana nature of the light neutrinos, which underlies processes such as neutrinoless double beta decay.

\subsection{Effective neutrino mass}\label{subsec:eff_nu_mass}
The Lagrangian in Eq.~(\ref{eq:mass_lagrangian}) is in the flavor basis. The mass matrix $\mathcal{M}$ in Eq.~(\ref{eq:mass_Matrix}) can be diagonalized, resulting in the diagonal matrix $\mathcal{D}$, which separates the masses into light and heavy neutrinos, as given by
\begin{equation}
    \mathcal{D} = \mathcal{U^T}\mathcal{M}\mathcal{U} = \begin{pmatrix}
        \mathcal{D_\nu} & 0 \\ 0 & \mathcal{D}_N
    \end{pmatrix}
\end{equation}
The diagonalization splits the masses into two block matrices $\mathcal{D_\nu}$ and $\mathcal{D}_N$. In the seesaw limit ($M_R\gg m_D$) $\mathcal{D_\nu}$ gives the active neutrino mass as $-m_DM_R^{-1}m_D^T$ and the sterile neutrino mass as $M_R$ (the diagonalization of the seesaw mass matrix is shown in Appendix~\ref{sec:mass_matrix_diagonalisation}). Using Euler's block parametrization, the mixing matrix $\mathcal{U}$ is parametrized as~\cite{Xing:2011ur}, 
\begin{equation}
    \mathcal{U} = \begin{pmatrix}
        \mathcal{I}&0\\0&U_{ss}
    \end{pmatrix}\begin{pmatrix}
        A&R\\S&B
    \end{pmatrix}\begin{pmatrix}
        U_0&0\\0&\mathcal{I}
    \end{pmatrix}\label{full_unitary_matrix}
\end{equation}
where $A, R, S$, and $B$ are block matrices of sterile-active mixing that give the interplay of active and sterile sectors with each other. Here, $U_0$ is the standard 3$\times$3 Pontecorvo-Maki-Nakagawa-Sakata (PMNS) matrix and $U_{ss}$ is the sterile-sterile mixing matrix. The explicit forms of the mixing matrices are shown in Appendix~\ref{sec:Mixing Matrices}. But we make one small assumption — to maximize the predictive power of the model, we make the simplifying assumption that there is no intrinsic mixing within the sterile sector, which is implemented by setting the sterile-sterile mixing matrix $U_{ss}=\mathcal{I}$. In this framework, any residual mixing in the sterile sector is not a free parameter but is directly induced by the active-sterile mixing required to generate the known neutrino oscillation parameters.
\subsection{Seesaw Relation}\label{subsec:seesaw_relation}
The requirement that $\mathcal{U}$ be unitary imposes conditions on the sterile-active mixing parameters such as $\mathcal{U}^\dagger \mathcal{U} = \mathcal{U} \mathcal{U}^\dagger = \mathcal{I}_{6}$. These conditions lead to relations between the block matrices of sterile-active mixing in the form of 
\begin{equation}
\begin{aligned}
    NN^\dagger + RR^\dagger = \mathcal{I}_3,\\
    TT^\dagger + BB^\dagger = \mathcal{I}_3,\\
    NT^\dagger + RB^\dagger = \mathcal{O}_3,
\end{aligned}\label{eq: uni_consequence}
\end{equation}
where $N=AU_0$ and $T=SU_0$ are the effective mixing matrix blocks of $\mathcal{U}$ from Eq.~(\ref{full_unitary_matrix}). $\mathcal{I}_3$ is the 3$\times$3 identity matrix and $\mathcal{O}_3$ is the null matrix of the same dimension. $N$ is the effective mixing matrix of the active-active sector. Since all matrices are finite (as sterile states do exist) and from Eq.~(\ref{eq: uni_consequence}), $N$ is non-unitary. This deviation is one of the key predictions of low-scale seesaw models and serves as a primary source for many phenomenological effects, such as charged lepton flavor violation~\cite{Flores-Tlalpa:2001vbz,Forero:2011pc}.
From the unitarity conditions in Eq.~(\ref{eq: uni_consequence}), by comparing the (1,1) block of $\mathcal{U}^T\mathcal{M}\mathcal{U} = \mathcal{D}$ (see Appendix~\ref{sec:seesaw_relation_appendix} for the full derivation), the exact seesaw relation between the active and sterile sectors follows as
\begin{equation}
\begin{aligned}
     U_0\mathcal{D_\nu}U_0^T = -(A^{-1}R)\mathcal{D}_N(A^{-1}R)^T
\end{aligned}\label{eq:Exact_Seesaw_relation}
\end{equation}
A detailed derivation of this is given in Appendix~\ref{sec:seesaw_relation_appendix}. This relation gives a mapping between low-scale parameters, which are the active neutrinos, and the high-scale parameters, which are the seesaw-induced sterile neutrinos. The LHS part of Eq.~(\ref{eq:Exact_Seesaw_relation}) is the light neutrino part, which is expressed in terms of active-neutrino mixing matrix $U_0$ and active neutrino mass matrix $\mathcal{D}_\nu$. The RHS part is the sterile-neutrino part, which is expressed in terms of active-sterile mixing matrix A and R, and the sterile neutrino mass matrix $\mathcal{D}_N$. This mapping between the active neutrino parameters and the sterile neutrino parameters allows us to establish a direct connection between them rather than treating them as independent. Using this mapping, we can derive standard oscillation parameters as a function of sterile neutrino parameters. The key practical advantage of this approach is that it allows us to systematically explore how variations in the sterile sector propagate into the observable quantities measured at oscillation experiments. To extract these observables, we build a Hermitian matrix using Eq.~(\ref{eq:Exact_Seesaw_relation}) as,
\begin{align}
H = (U_0\mathcal{D_\nu}U_0^T)(U_0\mathcal{D_\nu}U_0^T)^\dagger = R D_{N}^{} R^{T}_{} R^{*}_{} D_{N}^{}
R_{}^{\dagger}
\label{eq:herm_mat}
\end{align}
whose eigenvalues correspond to the light neutrino mass-squared values, while the associated eigenvectors determine the elements of the PMNS matrix. This procedure, following the formalism of Ref.~\cite{Xing:2024gmy}, provides a systematic framework for translating sterile-sector parameters into experimentally measurable quantities. The detailed derivation is presented in Appendix~\ref{sec:nu_mass_mix_appendix}.

The standard three-neutrino paradigm has $9$ defining parameters: 
\begin{itemize}
    \item[-] 3 Standard neutrino mass terms from $D_\nu$ = diag $m_1, m_2, m_3$. For Normal Hierarchy (NH), this makes $m_1$ as the lightest active neutrino mass and $m_3$ the heaviest active neutrino mass. In the case of Inverted Hierarchy (IH), the scenario changes as $m_2$ becomes the heaviest active neutrino mass and $m_3$ becomes the lightest active neutrino mass.  
    \item[-] 3 active-active mixing angles which are the standard neutrino mixing - $\theta_{12}, \theta_{13}, \theta_{23}$. Those are the mixing that define $U_0$ in Eq.~(\ref{full_unitary_matrix}).
    \item[-] 3 complex phases corresponding to the mixing angles -  $\delta_{12}, \delta_{13}, \delta_{23}$. By standard parametrization the $\delta_{13}$ phase is defined as $\delta_{CP}$. This is essentially because in neutrino oscillations, only the $\delta_{CP}$ phase survives to participate in the oscillations. 
\end{itemize}
Based on the oscillation experiments and detectors~\cite{DayaBay:2022orm,T2K:2021xwb,NOvA:2018gge,Esteban:2020cvm}, the parameter space of these standard parameters is quite stringent, with all mixing angles and mass-squared differences determined to percent-level precision, while $\delta_{CP}$ remains only loosely constrained. But with the addition of three sterile neutrino states, the parameter space has increased quite a bit. The seesaw mechanism adds a total of 21 new parameters into the framework: 
\begin{itemize}
    \item[-] $3$ sterile neutrino masses from $\mathcal{D}_N = (M_4, M_5, M_6)$. We consider the mass hierarchy to be consistent between the sterile neutrino masses as $M_4< M_5<M_6$.  
    \item[-] $9$ sterile-active mixing angles from the block matrices $A$ and $R$ (see Appendix~\ref{sec:Mixing Matrices}) of Eq.~(\ref{full_unitary_matrix}) ($\theta_{ij}$ for $i=1,2,3$ and $j=4,5,6$)
    \item[-] $9$ sterile phases for the respective sterile-active mixing angles ($\eta_{ij}$ for $i=1,2,3$ and $j=4,5,6$) 
\end{itemize}
The subject of interest remains how much these sterile parameters can contribute to the explanation of the phenomenological observations and help put better constraints on the standard neutrinos. To achieve this, we use the exact seesaw relation in Eq.~(\ref{eq:Exact_Seesaw_relation}) to span the parameter space of sterile neutrinos and constrain the resulting standard neutrino parameters under the constraints of Table~\ref{tab:constraints}.
\begin{table*}[!htbp]
\centering
\begin{tabular}{@{}l l l@{}}
\toprule
\textbf{Constraint Type} & \textbf{Experiment/Observation} & \textbf{Bound Used} \\
\midrule
\multirow{2}{*}{Mass-squared differences} & Solar oscillation ($\Delta m^2_{21}$) & $7.3 \times 10^{-5} < \Delta m^2_{21} < 7.71 \times 10^{-5}$ eV$^2$ \\
& Atmospheric oscillation ($\Delta m^2_{31}$) & $2.2 \times 10^{-3} < \Delta m^2_{31} < 2.7 \times 10^{-3}$ eV$^2$ \\
\midrule
\multirow{3}{*}{Mixing angles} & Solar mixing angle ($\theta_{12}$) & $31.38^\circ < \theta_{12} < 35.86^\circ$ \\
& Reactor mixing angle ($\theta_{13}$) & $8.20^\circ < \theta_{13} < 8.98^\circ$ \\
& Atmospheric mixing angle ($\theta_{23}$) & $40.3^\circ < \theta_{23} < 51.3^\circ$ \\
\midrule
CP-violating phase & Global oscillation data ($\delta_{CP}$) & $96^\circ < \delta_{CP} < 422^\circ$ \\
\midrule
\multirow{3}{*}{Cosmological bound} & Planck 2018 & $\sum m_i < 0.12$ eV \\
& DESI forecast & $\sum m_i < 0.072$ eV \\
& Future experiments & $\sum m_i < 0.06$ eV \\
\midrule
\multirow{3}{*}{Neutrinoless double beta decay} & KamLAND-Zen & $|m_{\beta\beta}| < 0.156$ eV \\
& GERDA & $|m_{\beta\beta}| < 0.079$ eV \\
\midrule
Charged lepton flavor violation & MEG experiment & $\mathcal{B}(\mu \to e\gamma) < 4.2 \times 10^{-13}$ \\
\bottomrule
\end{tabular}
\caption{Neutrino oscillation and theoretical constraints considered in the seesaw mechanism parameter space scan.}
\label{tab:constraints}
\end{table*}

\begin{figure*}[!htbp]
    \centering
    \includegraphics[width=\textwidth]{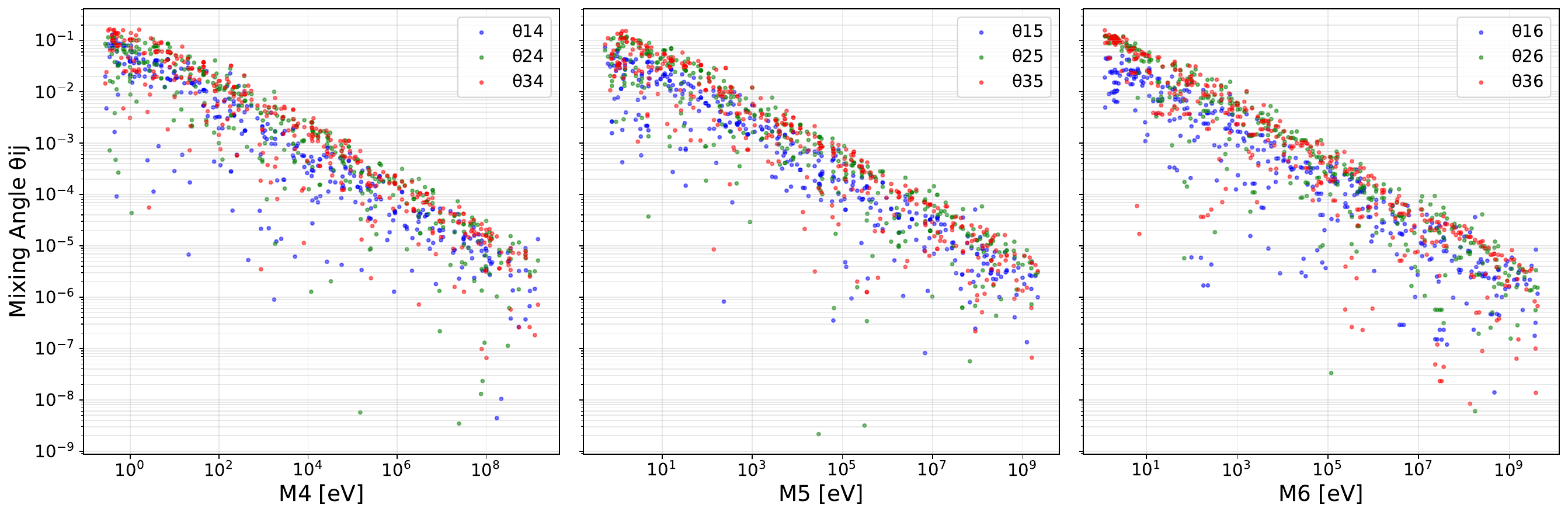}
    \caption{Viable seesaw parameter space in the sterile mass--mixing plane, obtained from a Monte Carlo scan of the 21-dimensional sterile parameter space subject to the oscillation and phenomenological constraints of Table~\ref{tab:constraints}. Each panel shows the lightest sterile neutrino mass $M_4$ versus the active--sterile mixing element $|\Theta_{\alpha 4}|^2$ for $\alpha = e,\mu,\tau$, respectively. The characteristic inverse-square scaling $M \propto \Theta^{-2}$ from the seesaw relation (Eq.~\ref{eq:Mass_vs_mixing}) is evident as a linear correlation on the logarithmic scale. Points are colour-coded by their sterile mass scale (eV, keV, MeV, GeV).}
    \label{fig:param_space_exploration}
\end{figure*}

The parameter space explored under these constraints can be seen in Fig.~(\ref{fig:param_space_exploration}). Under the bounds of the constraints, the exploration turns out to have a linear relation on a logarithmic scale. Using the exact seesaw relation in Eq.~(\ref{eq:Exact_Seesaw_relation}), we define the effective active--sterile mixing matrix as
\begin{align}
\Theta \equiv A^{-1}R .
\end{align}
Using this, Eq.~(\ref{eq:Exact_Seesaw_relation}) can be written as,
\begin{align}
D_\nu = - U_0^\dagger \, \Theta \, D_N \, \Theta^T \, U_0^* 
\end{align}
In the seesaw limit, this implies the characteristic scaling
$m_\nu \sim \Theta^2 M$ ,
With the active neutrino mass being the consistent parameter in the analysis, the relation can be interpreted as an inverse correlation between sterile mass and active-sterile mixing,
\begin{align}
M \propto \Theta^{-2}
\label{eq:Mass_vs_mixing}
\end{align}
In order to obtain the parameter space shown in Fig.~(\ref{fig:param_space_exploration}), we use the results obtained from solving Eq.~(\ref{eq:herm_mat}), which are explicitly derived in Appendix~\ref{sec:seesaw_relation_appendix}, to constrain the parameter space via Monte Carlo simulations of the full 21-dimensional sterile parameter space. Fig.~(\ref{fig:param_space_exploration}) is thus a visual representation of the viable parameter space obtained through these simulations. A selection of representative parameter sets is shown in Appendix~\ref{sec:seesaw_param_space}.
\subsection{Non-Unitarity effects}\label{subsec:non_unitarity}
One of the key consequences of the seesaw mechanism is the emergence of non-unitarity in the active neutrino mixing. In the standard $3$-flavor paradigm, the PMNS matrix $U_0$ is unitary, $U_0 U_0^\dagger = \mathcal{I}$, which guarantees the conservation of total oscillation probability among the three active flavors. The introduction of active-sterile mixing via the seesaw mechanism breaks this unitarity, as neutrino oscillations into sterile states become a finite possibility.

From the block decomposition in Eq.~(\ref{full_unitary_matrix}), the effective $3\times3$ mixing matrix governing active neutrino oscillations becomes $N = A \cdot U_0$, where $A$ is the lower-triangular non-unitary matrix encoding the active-sterile mixing. Since $A \neq \mathcal{I}$ in the presence of sterile neutrinos, the product $N$ is no longer unitary:
\begin{equation}\label{eq:non_unitarity_def}
    NN^\dagger = A U_0 U_0^\dagger A^\dagger = AA^\dagger \neq \mathcal{I}.
\end{equation}
The deviation from unitarity can be expressed element-wise through the active-sterile mixing block $R$ of the full $6\times6$ mixing matrix. Using the unitarity of the full matrix $\mathcal{U}$, we obtain
\begin{equation}\label{eq:non_unitarity_elements}
    (NN^\dagger)_{\alpha\beta} = \delta_{\alpha\beta} - \sum_{j=4}^{6} R_{\alpha j}\, R^*_{\beta j},
\end{equation}
where the sum runs over the three sterile mass eigenstates. The diagonal elements $|(NN^\dagger)_{\alpha\alpha} - 1| = \sum_{j=4}^{6} |R_{\alpha j}|^2$ directly measure the total active-sterile mixing for each active flavour $\alpha$, while the off-diagonal elements quantify lepton-flavour-violating correlations induced by the sterile sector.

\begin{figure}[!htb]
    \centering
    \includegraphics[width=\columnwidth]{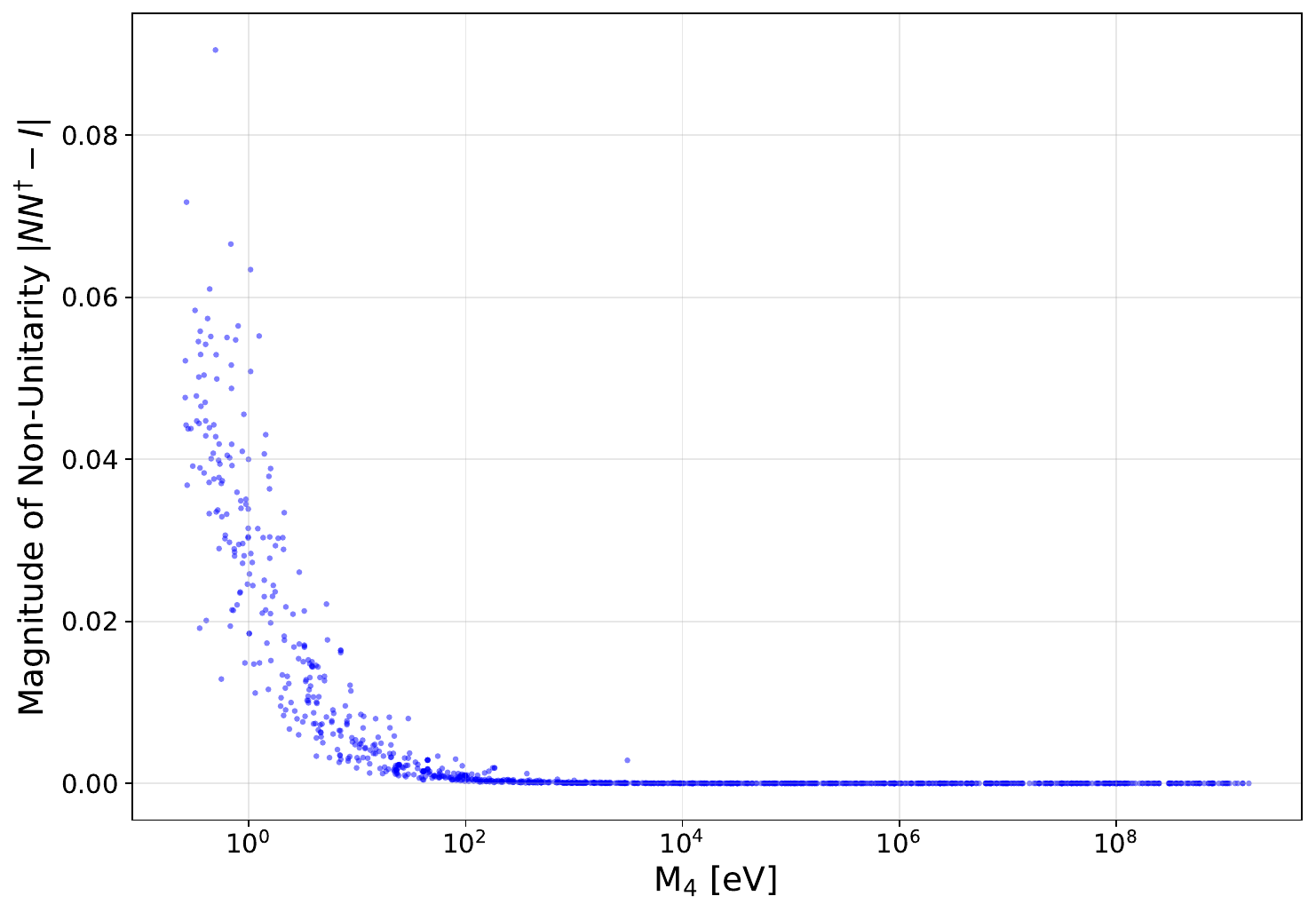}
    \caption{Non-unitarity of the active neutrino mixing matrix as a function of the lightest sterile neutrino mass $M_4$. The vertical axis shows $|NN^\dagger - \mathcal{I}|$, the Frobenius norm of the deviation of the effective $3\times 3$ mixing matrix from unitarity. The monotonic decrease reflects the seesaw-suppressed active-sterile mixing at higher mass scales (Eq.~\ref{eq:non_unitarity_elements}).}
    \label{fig:non_unitarity}
\end{figure}

We quantify the overall degree of non-unitarity through  $|NN^\dagger - \mathcal{I}|$, the Frobenius norm of the deviation matrix. From the seesaw relation, Eq.~(\ref{eq:Mass_vs_mixing}), the active-sterile mixing elements $R_{\alpha j}$ scale inversely with the sterile mass, $|R_{\alpha j}|^2 \propto M_j^{-1}$. Consequently, the non-unitarity monotonically decreases with increasing sterile mass scale, as illustrated in Fig.~\ref{fig:non_unitarity}. For eV-scale sterile neutrinos, the deviation can reach $\mathcal{O}(10^{-1})$, potentially observable at long-baseline experiments, whereas for GeV-scale and heavier sterile states, the non-unitarity is suppressed to $\mathcal{O}(10^{-5})$ or below, rendering it experimentally inaccessible with current facilities. This mass-dependent suppression is the underlying reason why heavier sterile neutrinos progressively decouple from oscillation experiments, a feature that will be demonstrated quantitatively in the next section.

\section{Analytical formulation of Oscillation probabilities within Seesaw framework}
\label{sec:seesaw_oscillations}
Neutrino oscillations 
in the standard three-neutrino paradigm, is the oscillation between the mass eigenstates $|\nu_i\rangle$ and the flavor eigenstates $|\nu_\alpha\rangle$, given by $|\nu_\alpha\rangle = \sum_i (U_0)_{\alpha i}|\nu_i\rangle$, where $U_0$ is the standard PMNS matrix defined in Eq.~(\ref{full_unitary_matrix}). The unitarity of the PMNS matrix ensures the conservation of the total oscillation probability~\cite{Bilenky:1987ty}.
The seesaw mechanism essentially adds sterile neutrino states to this picture. So, if we focus on the same neutrino oscillation probability and consider mixing between active states and sterile states, the mixing matrix extends to a 6$\times$6 unitary matrix given as 
\begin{equation}\label{eq:Unitary_matrix_representation}
\begin{aligned}
    \mathcal{U} = \;&O_{56}\,O_{46}\,O_{36}\,O_{26}\,O_{16}\\
    &\times O_{45}\,O_{35}\,O_{25}\,O_{15}\\
    &\times O_{34}\,O_{24}\,O_{14}\\
    &\times O_{23}\,O_{13}\,O_{12}
\end{aligned}
\end{equation}
where $O_{ij}$ are the rotation matrices for mixing of $i$ and $j$. The Euler block parametrization of this is what leads to Eq.~(\ref{full_unitary_matrix}). 
Using Eq.~(\ref{eq:Unitary_matrix_representation}), we define the neutrino oscillation probability as, 
\begin{equation}
\begin{split}
    P(\nu_\alpha \xrightarrow{}\nu_\beta) &= \delta_{\alpha \beta} \\
    &\quad - 4 \sum_{i>j}\Re{\mathcal{U}_{\alpha i}\mathcal{U}_{\beta i}^*\mathcal{U}_{\alpha j}^*\mathcal{U}_{\beta j}}\sin^2{(x_{ij})} \\
    &\quad + 2\sum_{i>j}\Im{\mathcal{U}_{\alpha i}\mathcal{U}_{\beta i}^*\mathcal{U}_{\alpha j}^*\mathcal{U}_{\beta j}}\sin{(2x_{ij})},
\end{split}
\end{equation}
In the $6\nu$ framework, $\mathcal{U}$ is the full $6\times6$ unitary mixing matrix.
Where $x_{ij} = \frac{\Delta m^2_{ij}L}{4E}$ is the oscillation phase, with $\Delta m^2_{ij} = m_i^2 - m_j^2$ being the mass-squared difference between the $i^{\mathrm{th}}$ and $j^{\mathrm{th}}$ neutrino mass eigenstates, $E$ the neutrino energy, and $L$ the propagation baseline. For $\alpha \text{ and }\beta$ being active neutrino flavors, i.e. $\{e,\mu,\tau\}$, this oscillation probability can be categorized into three parts — standard 3$\nu$ oscillations ($P_{3\nu}$), active -- sterile interference ($P_{sa}$) and sterile - sterile interference ($P_{ss}$) — as,
\begin{align}
    P_{3\nu} &= \delta_{\alpha \beta} - 4 \sum_{\substack{i>j \\ i,j=1}}^{3}\Re{(\mathcal{U}_{\alpha i}\mathcal{U}_{\beta i}^*\mathcal{U}_{\alpha j}^*\mathcal{U}_{\beta j})}\sin^2{(x_{ij})} \notag\\
             &\quad + 2\sum_{\substack{i>j \\ i,j=1}}^{3}\Im{(\mathcal{U}_{\alpha i}\mathcal{U}_{\beta i}^*\mathcal{U}_{\alpha j}^*\mathcal{U}_{\beta j})}\sin{(2x_{ij})} \label{eq:P3nu}\\
    P_{sa} &= - 4 \sum_{\substack{i=1,2,3 \\ j=4,5,6}}\Re{(\mathcal{U}_{\alpha i}\mathcal{U}_{\beta i}^*\mathcal{U}_{\alpha j}^*\mathcal{U}_{\beta j})}\sin^2{(x_{ij})} \notag\\
           &\quad + 2\sum_{\substack{i=1,2,3 \\ j=4,5,6}}\Im{(\mathcal{U}_{\alpha i}\mathcal{U}_{\beta i}^*\mathcal{U}_{\alpha j}^*\mathcal{U}_{\beta j})}\sin{(2x_{ij})} \label{eq:Psa}\\
    P_{ss} &= - 4 \sum_{\substack{i>j \\ i,j=4}}^{6}\Re{(\mathcal{U}_{\alpha i}\mathcal{U}_{\beta i}^*\mathcal{U}_{\alpha j}^*\mathcal{U}_{\beta j})}\sin^2{(x_{ij})} \notag\\
           &\quad + 2\sum_{\substack{i>j \\ i,j=4}}^{6}\Im{(\mathcal{U}_{\alpha i}\mathcal{U}_{\beta i}^*\mathcal{U}_{\alpha j}^*\mathcal{U}_{\beta j})}\sin{(2x_{ij})} \label{eq:Pss}
\end{align}
 The introduction of these $P_{sa}$ and $P_{ss}$ components into the framework are the result of adding the sterile states. These components, hence, affect the neutrino oscillations for the three active neutrinos. This is a direct consequence of the fact that the neutrino oscillation probability will be conserved.   
 
 Another point to note is that, although the sterile neutrinos do not interact with each other, there will still be a finite oscillation probability for the sterile neutrinos to oscillate into another sterile neutrino state. This can also be considered a form of probability leakage; however, it is effectively due to an existing yet extended route of sterile neutrinos oscillating into active flavors, and these active flavors subsequently oscillating into sterile states.

\begin{table*}[!htbp]
\centering
\renewcommand{\arraystretch}{1.2}
\begin{tabular}{|l|c|c|c|c|}
\hline\hline
\textbf{Parameter} 
& \textbf{eV scale} 
& \textbf{keV scale} 
& \textbf{MeV scale} 
& \textbf{GeV scale} \\
\hline
$\sin^2\theta_{12}$  & \multicolumn{4}{c|}{0.307} \\
$\sin^2\theta_{13}$  & \multicolumn{4}{c|}{0.0224} \\
$\sin^2\theta_{23}$  & \multicolumn{4}{c|}{0.546} \\
$\delta_{\rm CP}$ (deg) & \multicolumn{4}{c|}{212} \\
$\Delta m^2_{21}$ (eV$^2$) & \multicolumn{4}{c|}{$7.54\times10^{-5}$} \\
$\Delta m^2_{31}$ (eV$^2$) & \multicolumn{4}{c|}{$2.51\times10^{-3}$} \\
\hline
$\sin^2\theta_{sa}$ & $10^{-2}$ & $10^{-3}$ & $10^{-6}$ & $10^{-9}$ \\
$M_4$ (eV) & $10^{0}$ & $10^{3}$ & $10^{6}$ & $10^{9}$ \\
$M_5$ (eV) & $10^{0.2}$ & $10^{3.2}$ & $10^{6.2}$ & $10^{9.2}$ \\
$M_6$ (eV) & $10^{0.4}$ & $10^{3.4}$ & $10^{6.4}$ & $10^{9.4}$ \\
\hline\hline
\end{tabular}
\caption{Benchmark oscillation and sterile-sector parameters used in this work. 
The standard three-flavor oscillation parameters are fixed to their NuFIT~6.0 best-fit values for Normal Ordering \cite{Esteban:2024eli}, 
while sterile masses and active--sterile mixing angles are varied across different mass scales.}
\label{tab:parameters}
\end{table*}
 
In our framework, we assume no direct sterile-sterile mixing, rendering the $P_{ss}$ contribution negligible. For heavy sterile neutrinos, the oscillation phase $x_{ij} = \Delta m^2_{ij} L/(4E)$ becomes extremely large due to the large mass-squared splittings $\Delta m^2_{ij}$. Consequently, the rapid oscillations average out over the experimental energy resolution and baseline uncertainties. Under this averaging, $\langle \sin^2(x_{ij}) \rangle \to 1/2$ and $\langle \sin(2x_{ij}) \rangle \to 0$. The active-sterile transition probability $P_{sa}$ then reduces to its non-oscillatory component, becoming purely dependent on the active-sterile mixing angles with no observable oscillatory signatures. Combined with the seesaw-suppressed mixing for heavy sterile masses, the net effect of heavy sterile neutrinos becomes negligible at oscillation experiments.

So, in the higher mass regions of sterile neutrinos, this can be explicitly shown as 
\begin{equation}
    \langle P_{sa} \rangle_{\alpha\beta} = -2\sum_{\substack{i=1,2,3 \\ j=4,5,6}} \text{Re}\!\left(\mathcal{U}_{\alpha i}\,\mathcal{U}_{\beta i}^*\,\mathcal{U}_{\alpha j}^*\,\mathcal{U}_{\beta j}\right)
\end{equation} with $\langle \sin(2x_{ij}) \rangle$ becoming essentially zero. Now, using identity relations, we can write the mixing term as $\sum_{j=4}^{6}\mathcal{U}_{\alpha j}^*\,\mathcal{U}_{\beta j} = \delta_{\alpha\beta} - \sum_{i=1}^{3}\mathcal{U}_{\alpha i}^*\,\mathcal{U}_{\beta i} = \delta_{\alpha\beta} - (NN^\dagger)_{\alpha\beta}^*$. Using this, we obtain the expression of the active-sterile interference term as, 
\begin{equation}
    \langle P_{sa} \rangle_{\alpha\beta} = -2\,\text{Re}\!\left[(NN^\dagger)_{\alpha\beta}\left(\delta_{\alpha\beta} - (NN^\dagger)_{\beta\alpha}\right)\right]
\end{equation}
Keeping in mind that $N=AU_0$, the factors of $NN^\dagger  = AA^\dagger$ since $U_0$ are unitary, which will make the interference to be,
\begin{equation}
    \langle P_{sa} \rangle_{\alpha\beta} = -2\,\text{Re}\!\left[(AA^\dagger)_{\alpha\beta}\left(\delta_{\alpha\beta} - (AA^\dagger)_{\beta\alpha}\right)\right]
\end{equation}
This gives us the expression for an effective active-sterile interference in the active neutrino oscillations that is entirely dependent on the non-unitarity factor induced by the presence of sterile neutrinos. 
Similarly, we can also write the term for effective oscillation probability between the sterile states to be expressed as, 
\begin{align}
    \langle P_{ss} \rangle_{\alpha\beta} &= -2\sum_{\substack{k>l=1}}^{3}\text{Re}\!\left(R_{\alpha k}\,R_{\beta k}^*\,R_{\alpha l}^*\,R_{\beta l}\right) \notag\\
    &= |(RR^\dagger)_{\alpha\beta}|^2 - \sum_{k=1}^{3}|R_{\alpha k}|^2\,|R_{\beta k}|^2
\end{align}
So, the effect of coupling of sterile states to the active neutrinos arise entirely dependent on the active-sterile mixing. Now, with Eq.~(\ref{eq:Mass_vs_mixing}), the heavier neutrino states would effectively give very small mixing, leading to the suppression of the active-sterile mixing and thus the $P_{sa}$ and $P_{ss}$ factors being heavily suppressed or negligible.

\subsection{Long Baseline Experiments}\label{subsec:lbl_exp}
Long baseline neutrino oscillation experiments employ accelerator-produced neutrino beams and detectors positioned hundreds to thousands of kilometers from the source to probe neutrino mixing parameters and $CP$ violation through vacuum and matter-enhanced oscillations. We consider two benchmark experiments: NO$\nu$A (baseline $810$~km) and DUNE (baseline $1249$~km), which target precision measurements of $\Delta m^2_{31}$, $\theta_{23}$, $\delta_{CP}$, and mass ordering determination.

A critical consideration for LBL experiments is the matter effect~\cite{Wolfenstein:1977ue,Mikheyev:1985zog}, which arises from coherent forward scattering of neutrinos as they propagate through Earth's matter. Active neutrinos interact with electrons, protons, and neutrons via charged-current (CC) and neutral-current (NC) weak interactions, modifying the effective oscillation Hamiltonian. In contrast, sterile neutrinos are gauge singlets and do not experience these interactions. In the presence of sterile states, the modified Hamiltonian becomes
\begin{equation}
    H' = U^\dagger HU + V,
\end{equation}
where $V$ is the matter potential matrix of the form $\text{Diag}(V_{CC}+V_{NC},V_{NC},V_{NC},0,0,0)$. Here, $V_{CC}$ represents the charged-current interaction potential and $V_{NC}$ the neutral-current potential. Importantly, in the extended $6\nu$ framework, the $V_{NC}$ term, which cancels as a common phase in standard $3\nu$ oscillations, becomes physically relevant due to the differential matter interaction between active and sterile states.  
\subsubsection{Simulating $6\nu$ Oscillations}\label{subsec:6nu_sim}
The oscillation probabilities in the presence of sterile neutrinos depend sensitively on their masses and their mixing with the active sector. From Eq.~(\ref{eq:Mass_vs_mixing}), we observe that as sterile neutrino masses increase, the active–sterile mixing angles decrease correspondingly. Consequently, only sterile neutrinos in the lower mass range participate significantly in oscillation phenomena accessible to long-baseline experiments, while heavier sterile states become effectively decoupled from the active sector at these energies.

\begin{figure*}[!htbp]
    \centering
    \begin{subfigure}[t]{0.49\textwidth}
        \centering
        \includegraphics[width=\textwidth]{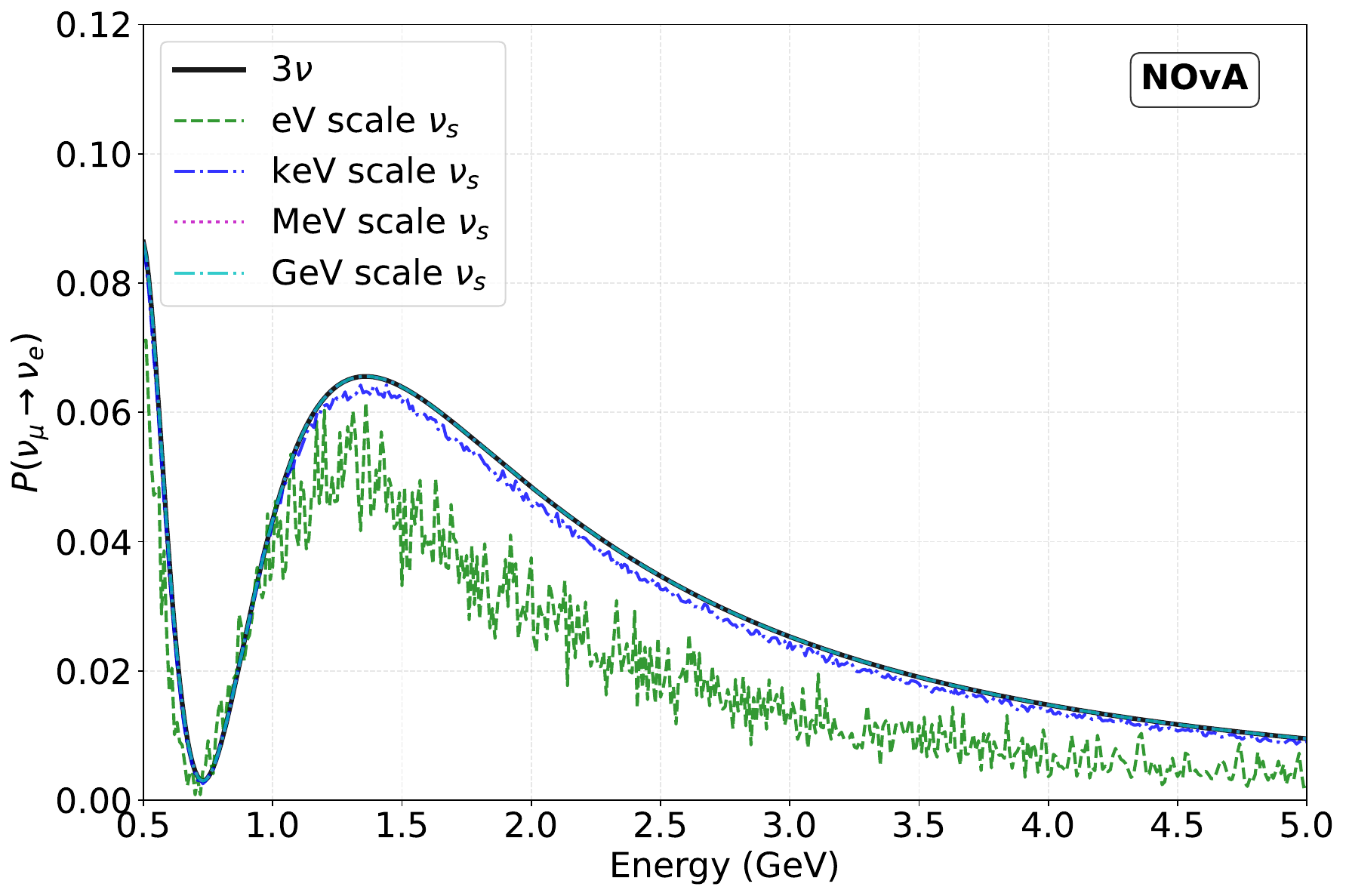}
        \caption{}
    \end{subfigure}
    \hfill
    \begin{subfigure}[t]{0.49\textwidth}
        \centering
        \includegraphics[width=\textwidth]{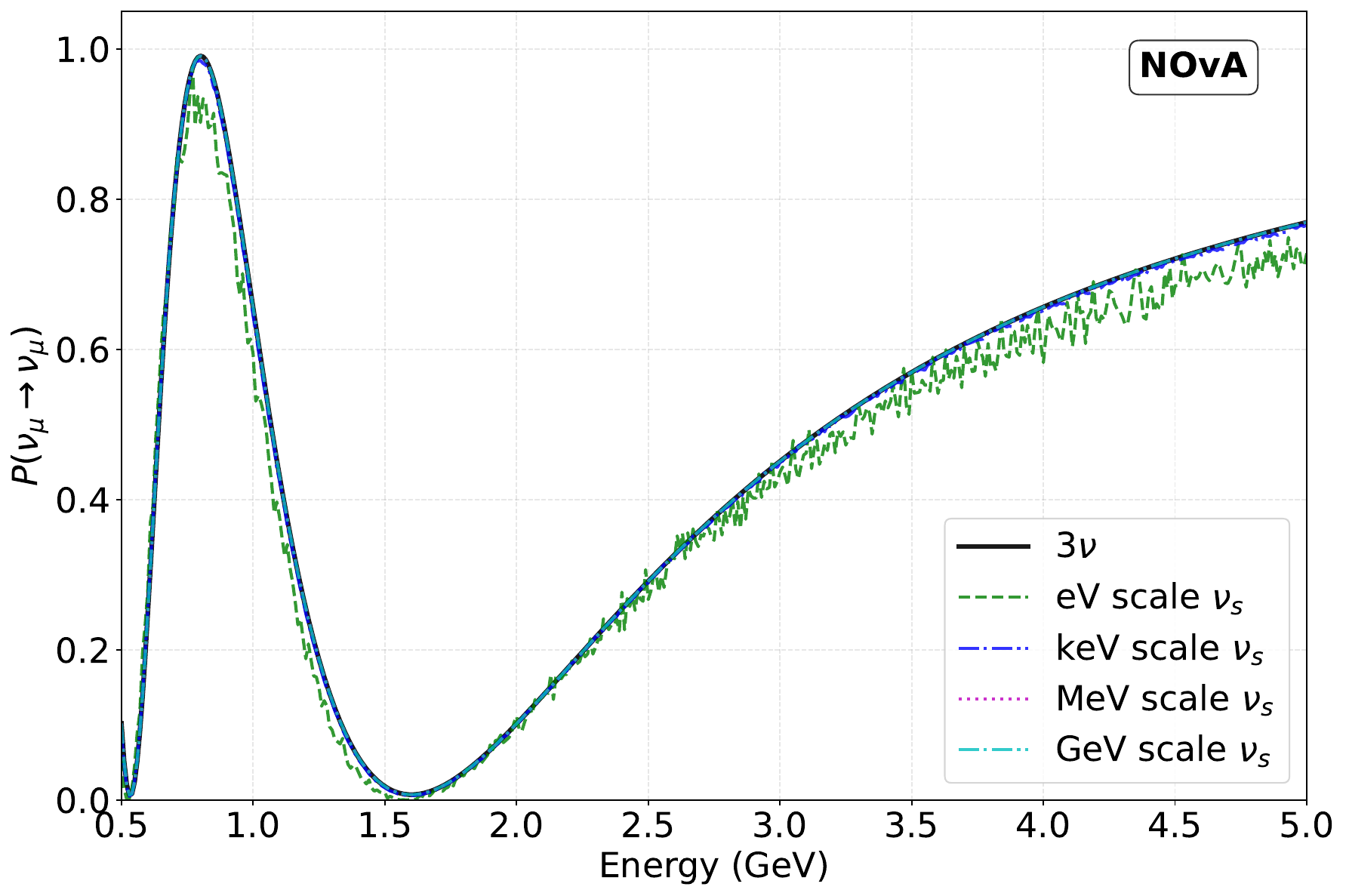}
        \caption{}
    \end{subfigure}
    \\[0.5em]
    \begin{subfigure}[t]{0.49\textwidth}
        \centering
        \includegraphics[width=\textwidth]{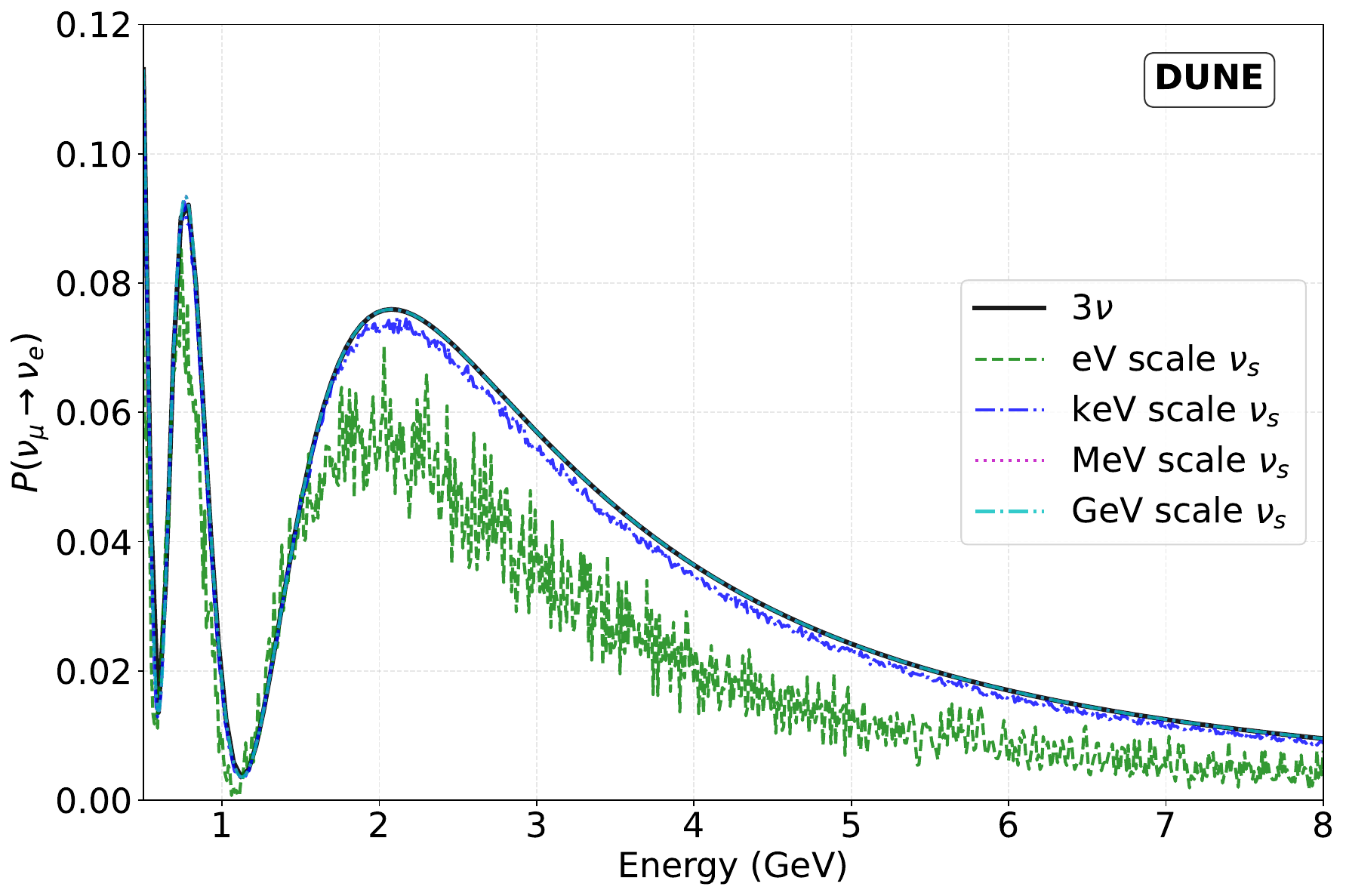}
        \caption{}
    \end{subfigure}
    \hfill
    \begin{subfigure}[t]{0.49\textwidth}
        \centering
        \includegraphics[width=\textwidth]{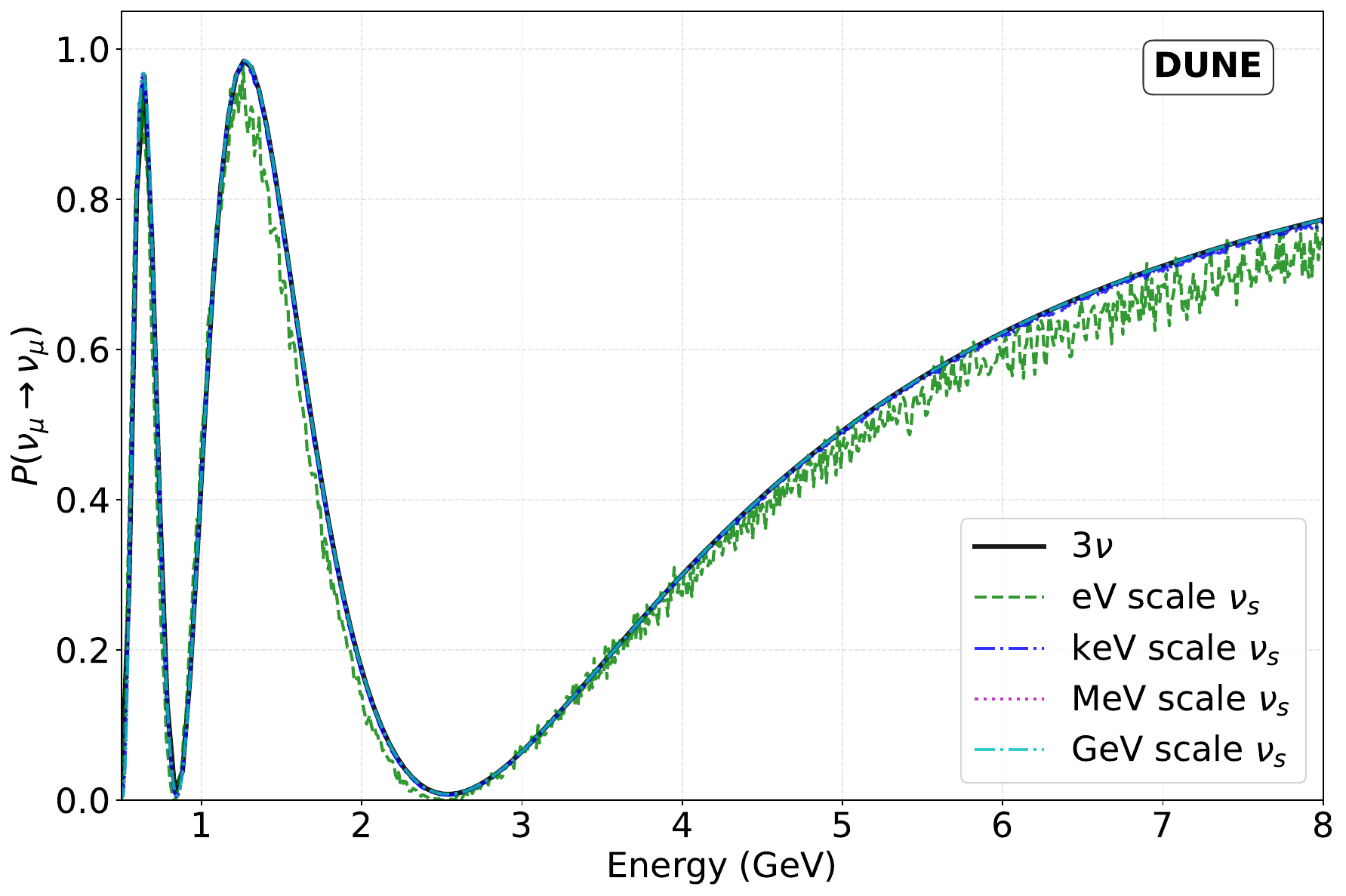}
        \caption{}
    \end{subfigure}
    \caption{Neutrino oscillation probabilities in the $6\nu$ framework: (a)~NO$\nu$A $\nu_\mu \to \nu_e$ appearance, (b)~NO$\nu$A $\nu_\mu \to \nu_\mu$ disappearance, (c)~DUNE $\nu_\mu \to \nu_e$ appearance, (d)~DUNE $\nu_\mu \to \nu_\mu$ disappearance. The standard three-flavor prediction (black) is compared against seesaw scenarios with sterile neutrinos at the eV (blue), keV (orange), MeV (green), and GeV (red) mass scales, using the benchmark parameters of Table~\ref{tab:parameters}. The eV-scale sterile neutrinos produce visible high-frequency oscillatory distortions, while heavier sterile states progressively decouple as the rapid oscillation phases average out.}
    \label{fig:Probability}
\end{figure*}

For sterile neutrinos at the eV mass scale, the associated mass-squared splittings induce oscillation phases comparable to those probed by experimental baselines and energies. Combined with finite active–sterile mixing, these states give rise to rapid, high-frequency oscillatory features in the transition probabilities, leading to visible deviations from the standard three-flavor oscillation pattern. This behavior is clearly illustrated in Fig.~(\ref{fig:Probability}), where eV-scale sterile neutrinos produce pronounced distortions relative to the standard three-flavor scenario.

In contrast, for heavier sterile neutrinos in the keV, MeV, and GeV mass ranges, the active–sterile mixing angles are strongly suppressed according to the seesaw scaling relation. Additionally, the corresponding oscillation phases vary extremely rapidly over the experimental energy resolution and baseline spread. Consequently, the associated oscillatory terms average out over detector acceptances, rendering these heavy sterile states effectively invisible to oscillation experiments. Their net effect reduces to an overall suppression of the active-sector probabilities, mimicking standard three-flavor oscillations with a reduced effective normalization.

This loss of sensitivity to heavy sterile neutrinos can be interpreted as an effective decoherence arising from the averaging of fast oscillations over experimental resolutions, rather than from fundamental environmental decoherence mechanisms. As a result, only sterile neutrinos with sufficiently small masses (eV to keV scale) and appreciable mixing can leave observable imprints in long-baseline oscillation measurements. 

\begin{figure*}[!htbp]
    \centering
    \begin{subfigure}[t]{0.49\textwidth}
        \centering
        \includegraphics[width=\textwidth]{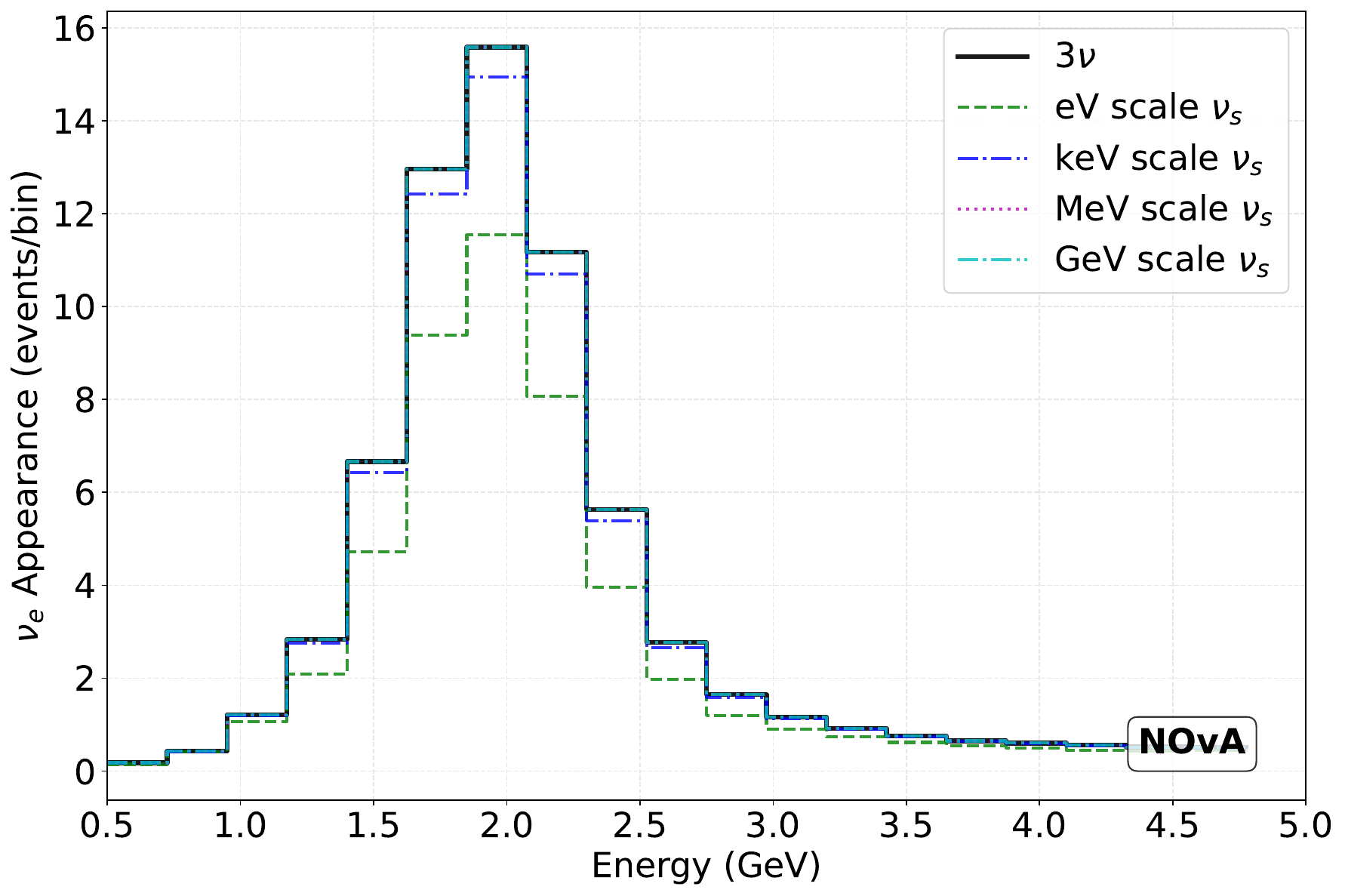}
        \caption{}
    \end{subfigure}
    \hfill
    \begin{subfigure}[t]{0.49\textwidth}
        \centering
        \includegraphics[width=\textwidth]{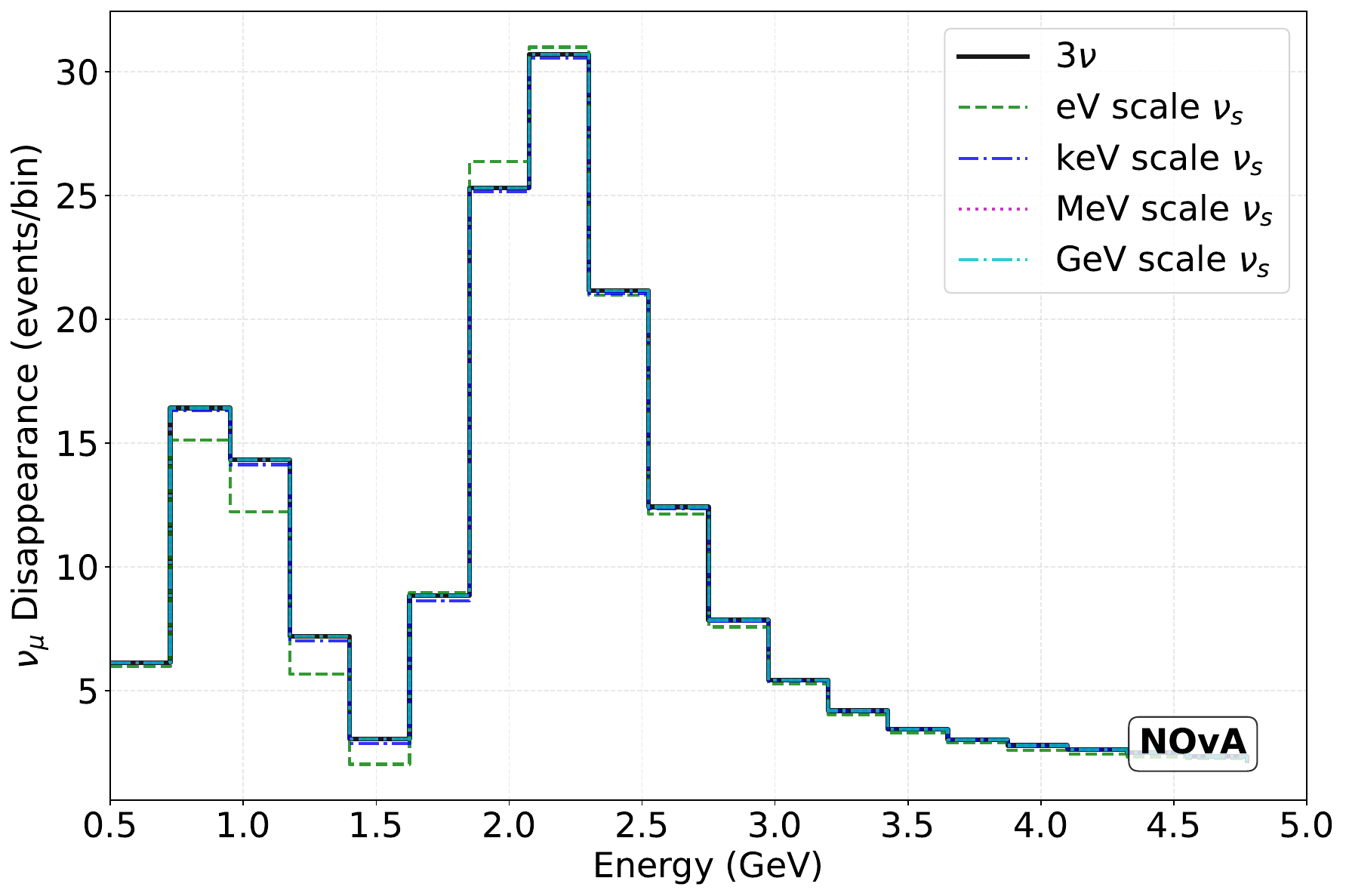}
        \caption{}
    \end{subfigure}
    \\[0.5em]
    \begin{subfigure}[t]{0.49\textwidth}
        \centering
        \includegraphics[width=\textwidth]{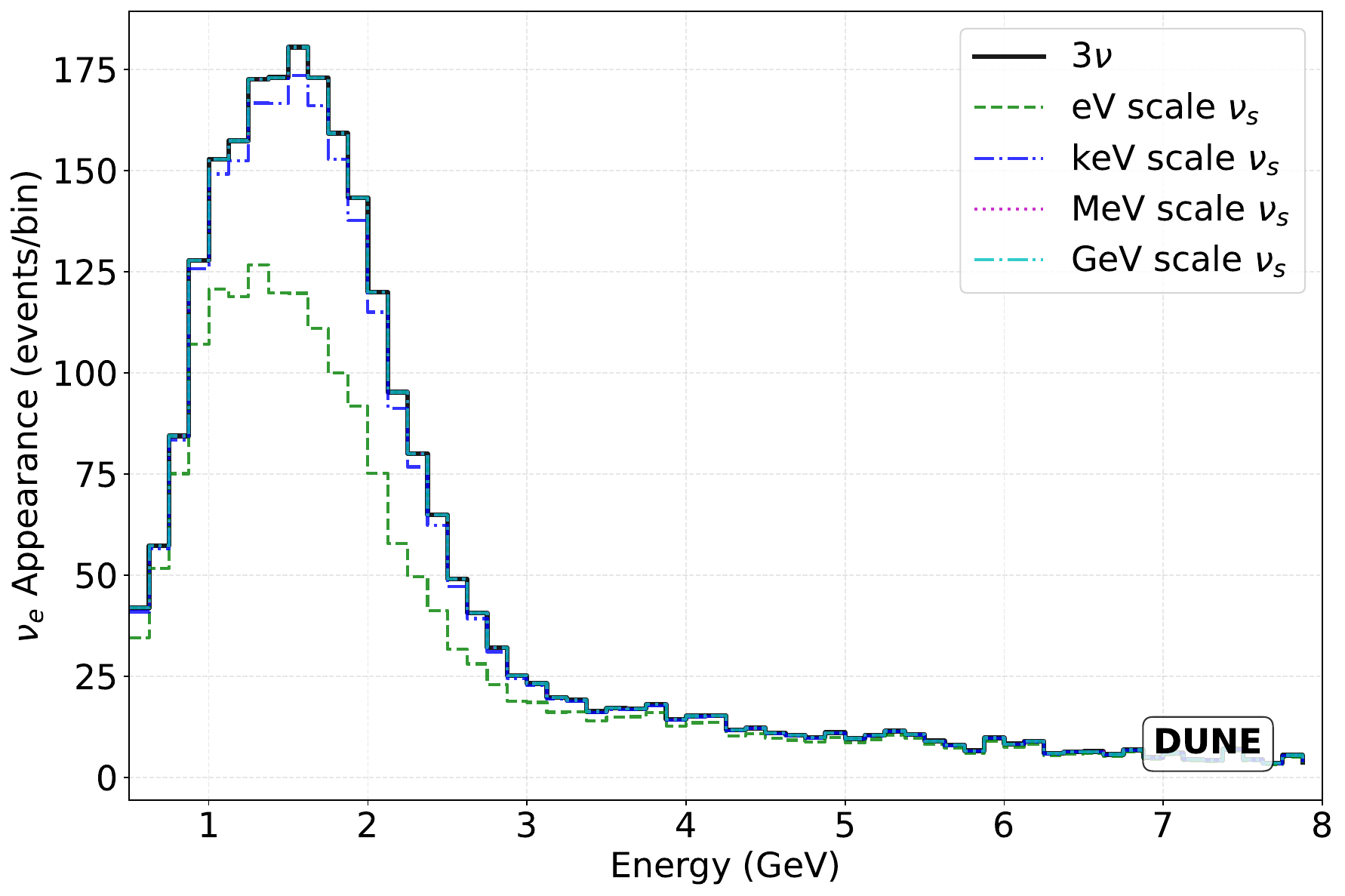}
        \caption{}
    \end{subfigure}
    \hfill
    \begin{subfigure}[t]{0.49\textwidth}
        \centering
        \includegraphics[width=\textwidth]{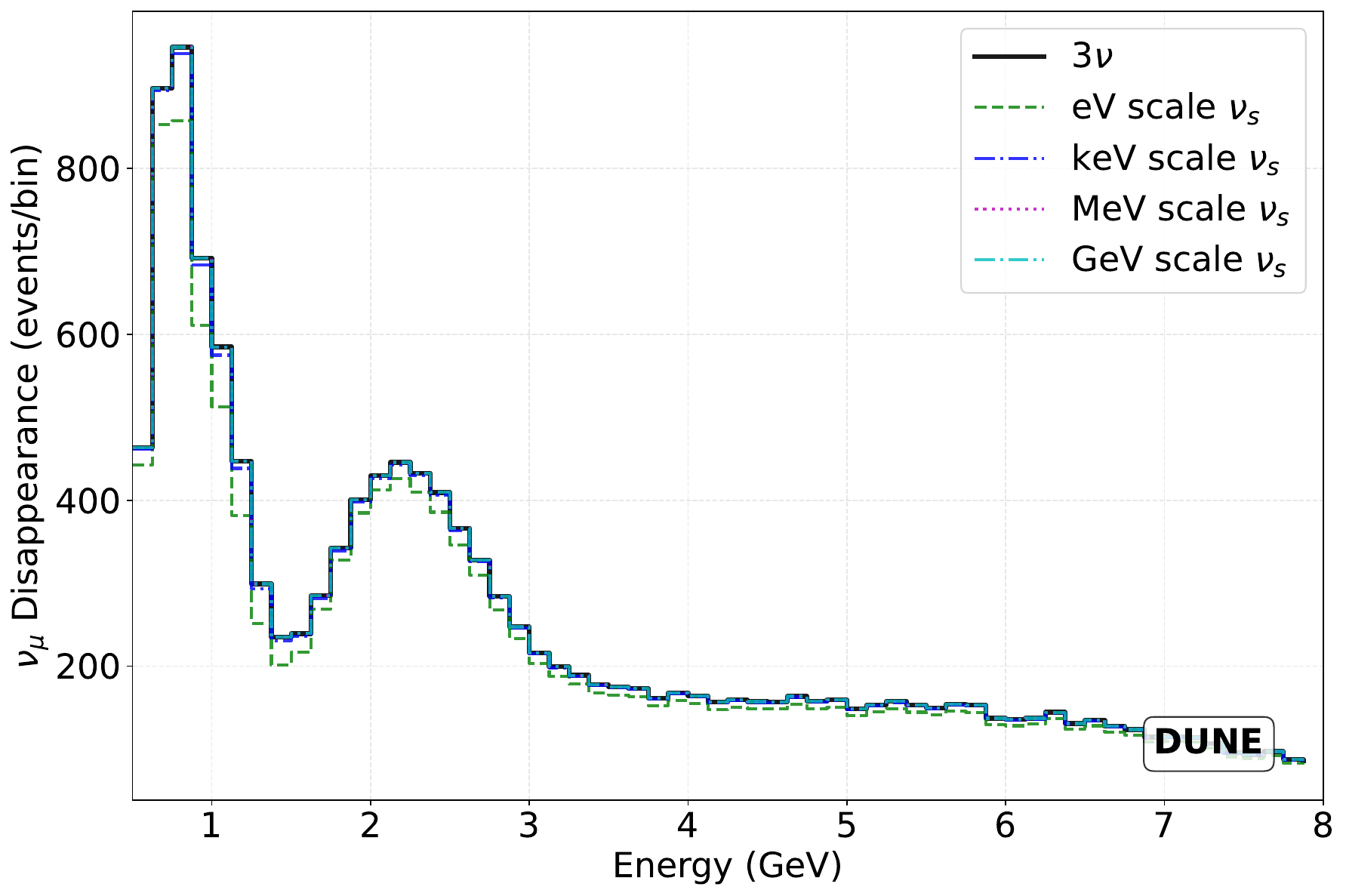}
        \caption{}
    \end{subfigure}
    \caption{Simulated event rates: (a)~NO$\nu$A $\nu_e$ appearance, (b)~NO$\nu$A $\nu_\mu$ disappearance, (c)~DUNE $\nu_e$ appearance, (d)~DUNE $\nu_\mu$ disappearance, including detector efficiencies and exposure. Each curve corresponds to a different sterile mass scale as in Fig.~\ref{fig:Probability}. The eV-scale sterile neutrinos induce measurable distortions in the event spectra, whereas keV--GeV-scale scenarios produce rates nearly indistinguishable from the standard three-flavor expectation due to the seesaw-suppressed mixing.}
    \label{fig:event_rates}
\end{figure*}

To quantify the experimental sensitivity to sterile neutrino effects, we simulate event rates for both NO$\nu$A and DUNE. Long baseline experiments primarily detect $\nu_\mu$ ($\overline{\nu_\mu}$) and $\nu_e$ ($\overline{\nu_e}$) events in FHC (RHC) mode, respectively. Therefore, all observable effects of sterile neutrinos must manifest through modifications to these detection channels. We focus on two key oscillation channels: $\nu_e$ appearance (sensitive to $\nu_\mu \to \nu_e$ transitions) and $\nu_\mu$ disappearance (sensitive to $\nu_\mu$ survival probability). As shown in Fig.~(\ref{fig:event_rates}), the behavior observed in the oscillation probability calculations translates directly to the predicted event rates, with eV-scale sterile neutrinos producing the most significant deviations from the standard three-flavor expectations.

\subsubsection{CP-Phase Effects on $6\nu$ Oscillations}\label{subsec:cpv_effects}
The oscillation probabilities show strong dependence on the CP-violating phase $\delta_{\rm CP}$. To make this dependence explicit, we recall that in the standard three-flavor framework the $\nu_\mu \to \nu_e$ appearance probability in vacuum can be decomposed as
\begin{equation}\label{eq:Pmue_decomposition}
P(\nu_\mu \to \nu_e) \simeq P_{\rm atm} + P_{\rm sol} + P_{\rm int}^{\cos\delta} + P_{\rm int}^{\sin\delta},
\end{equation}
where the individual contributions are
\begin{align}
P_{\rm atm} &= \sin^2\theta_{23}\,\sin^2 2\theta_{13}\,\sin^2\Delta_{31}, \label{eq:Patm}\\
P_{\rm sol} &= \cos^2\theta_{23}\,\sin^2 2\theta_{12}\,\sin^2\Delta_{21}, \label{eq:Psol}\\
P_{\rm int}^{\cos\delta} &= J_r\,\cos\delta_{\rm CP}\,\cos\Delta_{31}\,\sin\Delta_{21}\,\sin\Delta_{31}, \label{eq:Pcos}\\
P_{\rm int}^{\sin\delta} &= - J_r\,\sin\delta_{\rm CP}\,\sin\Delta_{31}\,\sin\Delta_{21}\,\sin\Delta_{31}, \label{eq:Psin}
\end{align}
with $\Delta_{ij} \equiv \Delta m^2_{ij}L/(4E)$ and $J_r = 8\,\cos\theta_{13}\,\sin 2\theta_{12}\,\sin 2\theta_{13}\,\sin 2\theta_{23}$ being the reduced Jarlskog invariant. The CP-odd term $P_{\rm int}^{\sin\delta}$ is the sole contribution that changes sign under CP conjugation ($\nu \leftrightarrow \bar\nu$) and under $\delta_{\rm CP} \to -\delta_{\rm CP}$. In the upper half-plane ($0 < \delta_{\rm CP} < \pi$), $\sin\delta_{\rm CP} > 0$, rendering $P_{\rm int}^{\sin\delta}$ negative for neutrinos and positive for antineutrinos with the overall sign governed by $\sin\Delta_{31}$. Combined with the CP-even interference term $P_{\rm int}^{\cos\delta}$, the net effect at LBL baselines is a constructive enhancement of the $\nu_e$ appearance signal relative to the lower half-plane, where the interference is predominantly destructive.

In the $6\nu$ seesaw framework, the appearance probability acquires additional CP-violating contributions from the active--sterile sector. Using the decomposition of Eq.~(\ref{eq:Psa}), the active--sterile interference introduces terms of the form
\begin{equation}\label{eq:Psa_CP}
P_{sa}^{\rm CP} = 2\sum_{\substack{i=1,2,3 \\ j=4,5,6}} \text{Im}\!\left(\mathcal{U}_{\mu i}\,\mathcal{U}_{e i}^*\,\mathcal{U}_{\mu j}^*\,\mathcal{U}_{e j}\right)\sin(2x_{ij}),
\end{equation}
where the mixing matrix elements $\mathcal{U}_{\alpha j}$ for $j = 4,5,6$ depend on the nine sterile phases $\eta_{ij}$. Crucially, these new imaginary parts do not vanish even when $\delta_{\rm CP} = 0$, implying that sterile neutrinos can generate CP-violating effects in oscillation experiments purely through the active--sterile mixing phases, independently of the standard Dirac phase. For eV-scale sterile states, the oscillation phases $x_{ij}$ remain resolvable and these additional CP-odd terms can produce observable modifications to the appearance signal. For heavier sterile masses, the rapid oscillations average out ($\langle\sin(2x_{ij})\rangle \to 0$), suppressing the sterile CP-odd contribution and leaving only the standard $\delta_{\rm CP}$-dependent terms.

The $\nu_\mu$ disappearance channel, in contrast, is largely insensitive to $\delta_{\rm CP}$ at leading order. This follows directly from the structure of the survival probability: for $\alpha = \beta$, the product $\mathcal{U}_{\alpha i}\,\mathcal{U}_{\alpha i}^*\,\mathcal{U}_{\alpha j}^*\,\mathcal{U}_{\alpha j} = |\mathcal{U}_{\alpha i}|^2\,|\mathcal{U}_{\alpha j}|^2$ is manifestly real, so the CP-odd imaginary part in Eq.~(\ref{eq:P3nu}) vanishes identically:
\begin{equation}
\text{Im}\!\left(|\mathcal{U}_{\mu i}|^2\,|\mathcal{U}_{\mu j}|^2\right) = 0 \quad \forall\; i,j.
\end{equation}
Consequently, $P(\nu_\mu \to \nu_\mu)$ depends only on the moduli $|\mathcal{U}_{\mu i}|^2$ and is independent of all CP phases at leading order. The corresponding event rates, therefore, remain essentially unchanged across different values of $\delta_{\rm CP}$.

\begin{figure*}[!htbp]
    \centering
    \begin{subfigure}[t]{0.49\textwidth}
        \centering
        \includegraphics[width=\textwidth]{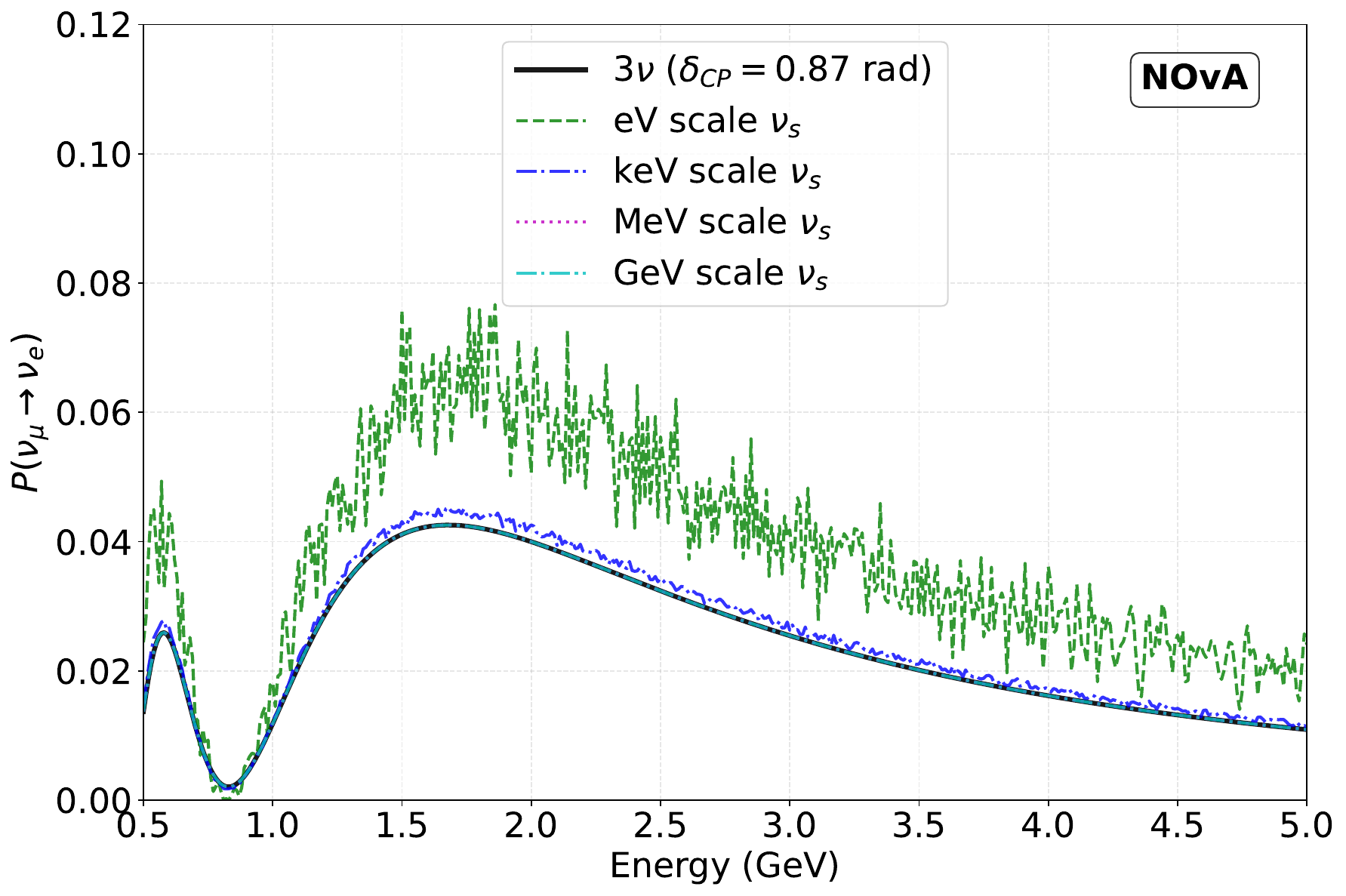}
        \caption{}
    \end{subfigure}
    \hfill
    \begin{subfigure}[t]{0.49\textwidth}
        \centering
        \includegraphics[width=\textwidth]{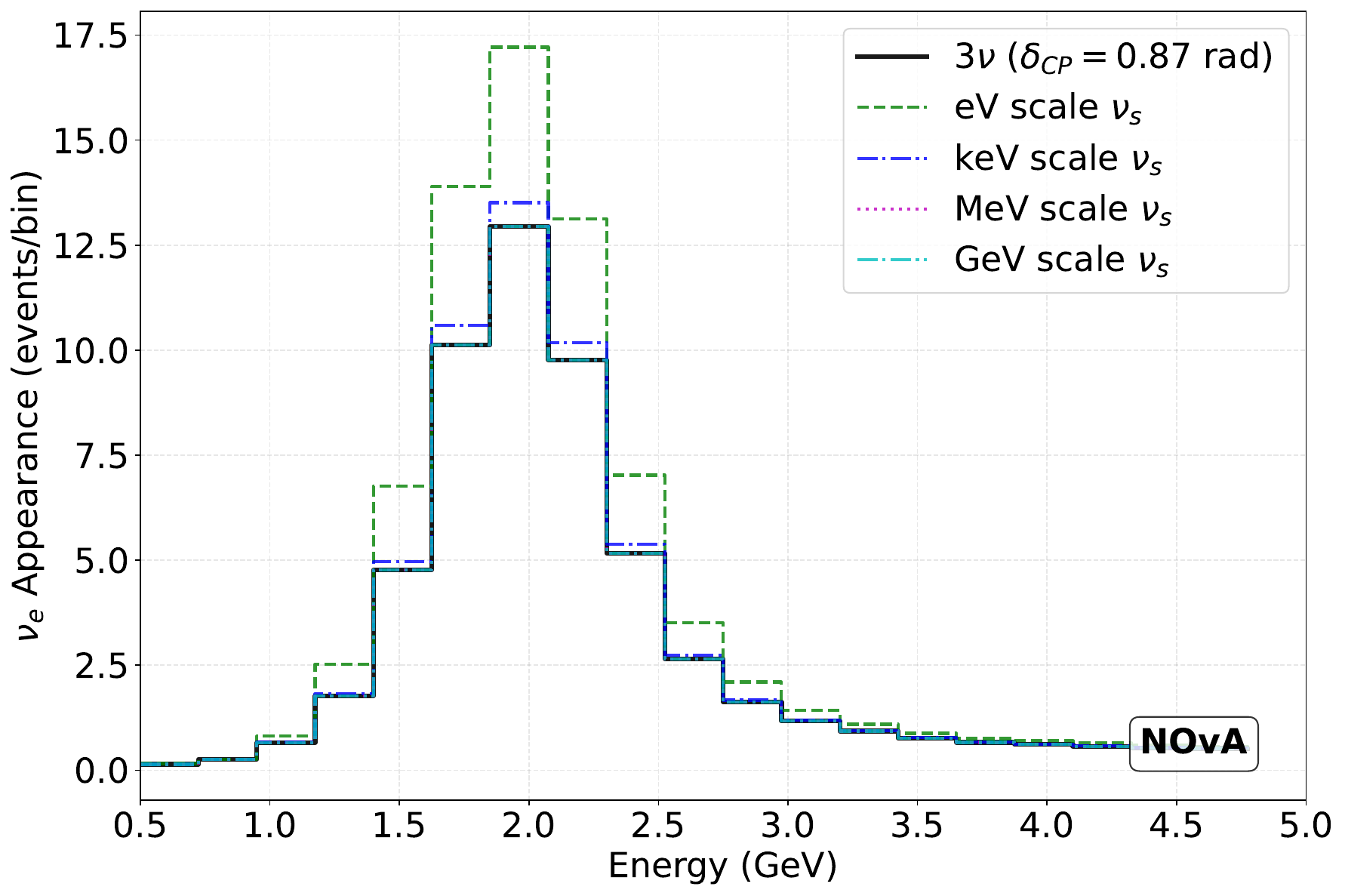}
        \caption{}
    \end{subfigure}
    \caption{Effect of the CP-violating phase on sterile neutrino signatures at NO$\nu$A. (a)~$\nu_\mu \to \nu_e$ appearance probability; (b)~corresponding $\nu_e$ appearance event rates, both evaluated at $\delta_{\rm CP} = 0.87$~rad (upper half-plane). Compared with the lower half-plane value used in Fig.~\ref{fig:Probability}, the constructive CP-dependent interference enhances the appearance signal. The eV-scale sterile neutrino distortions (blue) remain clearly distinguishable from the three-flavor prediction (black) in both CP-phase configurations.}
    \label{fig:nova_oscillations_dcp087}
\end{figure*}

For all simulations presented in this work, the oscillation and sterile-sector parameters are fixed to the values listed in Table~\ref{tab:parameters}. Despite extensive experimental efforts, no stringent constraints on $\delta_{\rm CP}$ have yet been established. For a representative comparison, we perform simulations of oscillation probabilities and event rates for $\delta_{\rm CP}$ in the upper half-plane, using parameter values consistent with the NOvA analysis~\cite{NOvA:2025tmb}. The resulting constructive enhancement of the appearance signal is illustrated in Fig.~(\ref{fig:nova_oscillations_dcp087}), where the eV-scale sterile neutrino distortions remain clearly distinguishable from the three-flavor prediction in both CP-phase configurations.

Although several global analyses exhibit a preference for values of $\delta_{\rm CP}$ in the lower half-plane \cite{T2K:2025wet}, no definitive constraints have yet been established.

\subsection{JUNO Simulation}\label{subsec:juno_sim}

The Jiangmen Underground Neutrino Observatory (JUNO) is a medium-baseline reactor neutrino experiment located in southern China. With a baseline of approximately 52.5~km from multiple nuclear reactor cores and a massive 20-kiloton liquid scintillator detector, JUNO is designed to achieve unprecedented energy resolution of 3\% at 1~MeV \cite{JUNO:2015zny}. The primary physics goals of JUNO include the determination of the neutrino mass ordering through precision measurements of the solar mass-squared splitting $\Delta m^2_{21}$ and the atmospheric splitting $\Delta m^2_{31}$, as well as high-precision measurements of the solar mixing angle $\theta_{12}$. Additionally, JUNO will have enhanced sensitivity to astrophysical neutrinos, including supernova neutrinos, solar neutrinos, and geoneutrinos.

For our analysis, we simulate a JUNO-like medium-baseline reactor antineutrino experiment using a reactor flux spectrum and baseline configuration consistent with JUNO's operating conditions. While we capture the essential oscillation physics relevant for JUNO—particularly its sensitivity to spectral distortions in the few-MeV energy range—we do not attempt to reproduce the full detector-specific response or all systematic effects. The resulting event rates therefore provide a phenomenological estimate of JUNO-like sensitivities suitable for studying sterile neutrino effects in this energy and baseline regime.

JUNO operates at MeV energies, which enables unique sensitivity to oscillation effects that are suppressed or averaged out in GeV-scale long-baseline experiments. We focus on the dominant detection channel: $\overline{\nu_e}$ disappearance (i.e., $\overline{\nu_e}\to\overline{\nu_e}$ survival probability). 

\begin{figure*}[!htbp]
    \centering
    \begin{subfigure}[t]{0.48\textwidth}
        \centering
        \includegraphics[width=\textwidth]{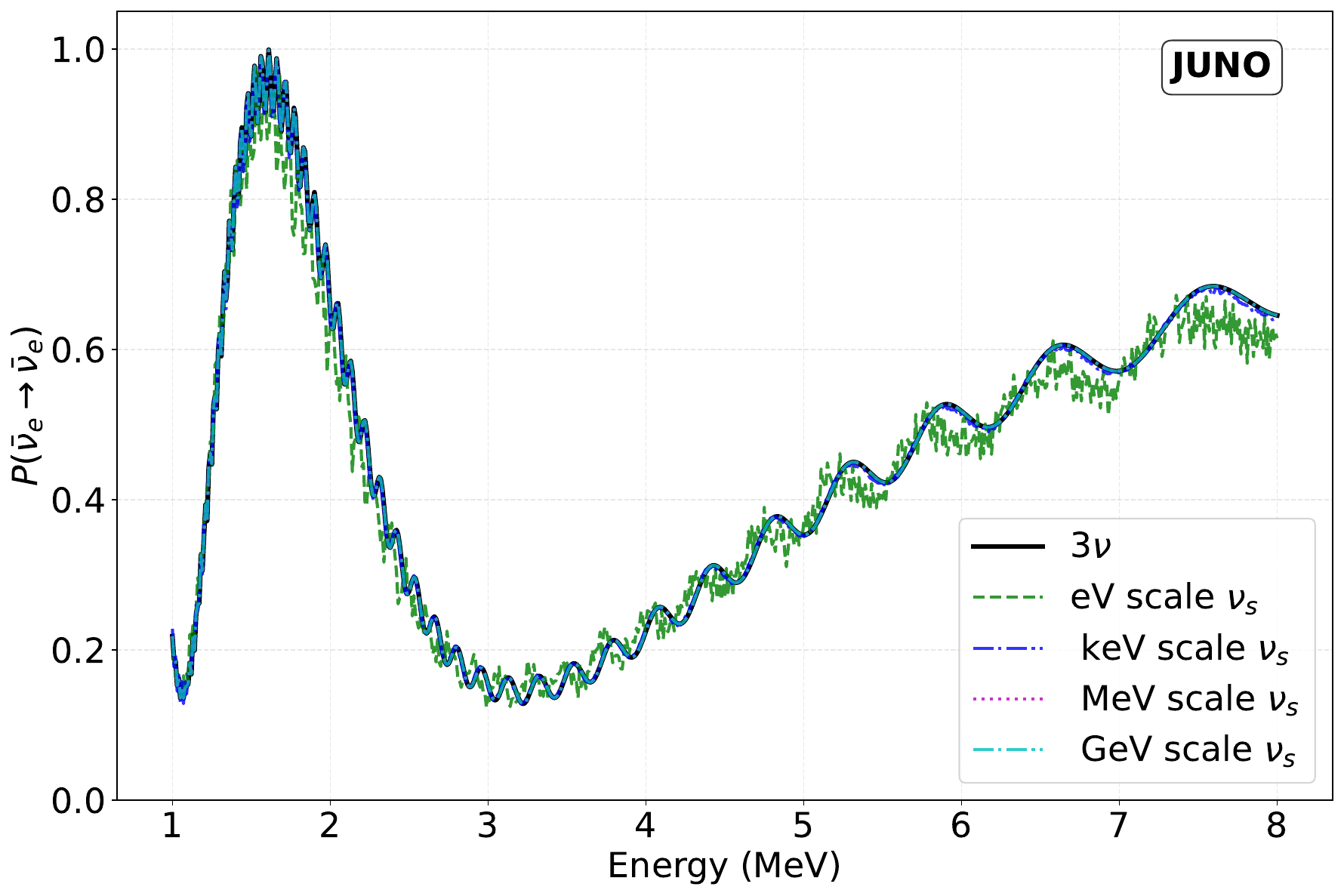}
        \caption{}
        \label{fig:juno_probability}
    \end{subfigure}
    \hfill
    \begin{subfigure}[t]{0.48\textwidth}
        \centering
        \includegraphics[width=\textwidth]{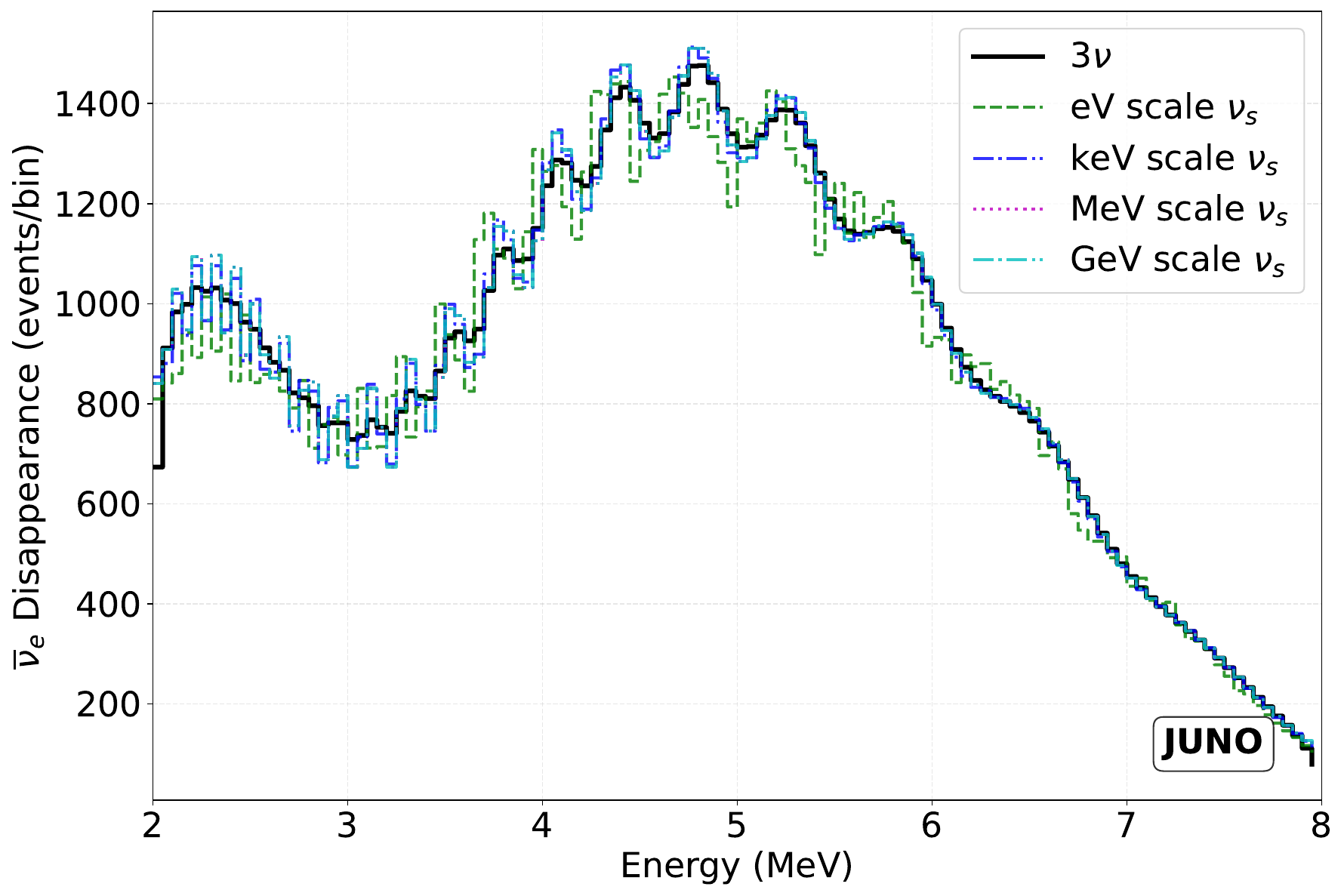}
        \caption{}
        \label{fig:juno_event_rates}
    \end{subfigure}
    \caption{JUNO sensitivity to sterile neutrino effects. (a)~Reactor $\overline{\nu_e} \to \overline{\nu_e}$ survival probability at the JUNO baseline of 52.5~km, for the standard three-flavor case (black) and the $6\nu$ seesaw scenarios at the eV (blue), keV (orange), MeV (green), and GeV (red) mass scales. eV-scale sterile neutrinos introduce pronounced high-frequency modulations, while keV- and MeV-scale states produce subtler but resolvable distortions. (b)~Simulated $\overline{\nu_e}$ disappearance event rates at JUNO for the same scenarios. The high statistics and excellent 3\% energy resolution of JUNO translate the probability-level distortions into experimentally distinguishable spectral features.}
    \label{fig:juno_combined}
\end{figure*}

In long-baseline experiments operating at GeV energies, the oscillation phases associated with $\Delta m^2_{31}$ and $\Delta m^2_{32}$ are nearly degenerate, while the solar phase driven by $\Delta m^2_{21}$ remains negligibly small. In contrast, for medium-baseline reactor experiments such as JUNO operating at MeV energies, the solar oscillation phase becomes $\mathcal{O}(1)$, leading to a distinct interference pattern between atmospheric-scale and solar-scale oscillations. This produces characteristic ``wiggle'' structures in the energy spectrum, as shown in Fig.~\ref{fig:juno_probability}. 

Similar to the behavior observed in long-baseline experiments, eV-scale sterile neutrinos induce rapid, high-frequency oscillatory features in the survival probability. However, unlike the LBL case, even heavier sterile neutrinos at the keV and MeV scales show non-negligible deviations from the standard three-flavor expectation at JUNO. While these deviations appear modest in the oscillation probability itself, they translate into significant effects on the observed event rates due to JUNO's excellent energy resolution and high statistics, as illustrated in Fig.~\ref{fig:juno_event_rates}.

\subsection{Sensitivity Analysis}\label{subsec:sensitivity_analysis}
The sensitivity of neutrino oscillation experiments to sterile neutrino effects is determined by their ability to distinguish sterile-induced spectral distortions from the standard three-flavor oscillation predictions. To quantify this sensitivity, we perform a $\chi^2$ analysis comparing predicted event rates in the presence of sterile neutrinos against the three-flavor benchmark. We employ a Poissonian $\chi^2$ statistic given by
\begin{equation}
\chi^2 = \sum_i \left( \mu_i - n_i + n_i \ln \frac{n_i}{\mu_i} \right)
+ \sum_i \left( \frac{x_i - x_{i,\mathrm{center}}}{\sigma_i} \right)^2,
\end{equation}
where $n_i$ denotes the number of observed events in the $i^{\mathrm{th}}$ energy bin, $\mu_i$ represents the corresponding predicted number of events under the sterile neutrino hypothesis, and the second term accounts for systematic uncertainties through nuisance parameters $x_j$ with central values $x_{j,\mathrm{center}}$ and associated one-sigma uncertainties $\sigma_j$.

The systematic uncertainties encapsulate various experimental effects, including signal and background normalization uncertainties, detection efficiencies, energy calibration errors, and background contamination for each oscillation channel. These nuisance parameters are profiled (marginalized) over in the $\chi^2$ minimization to obtain the sensitivity reach of each experiment.

We simulate NO$\nu$A (baseline 810~km) with exposures of $26.60\times10^{20}$ POT in FHC mode (6 years) and $12.50\times10^{20}$ POT in RHC mode (3 years), assuming systematic uncertainties of 5\% on signal normalization, 2.5\% on signal energy calibration, 10\% on background normalization, and 2.5\% on background calibration. For DUNE (baseline 1249~km), we adopt $71.5\times10^{20}$ POT distributed equally between FHC and RHC modes over 6 years each, with systematic uncertainties of 2\% on $\nu_e$ and $\overline{\nu_e}$ signal events, 5\% on $\nu_\mu$ and $\overline{\nu_\mu}$ signal events, 5\% on beam-related background events, and 10\% on neutral-current background events. For JUNO (baseline 52.5~km), we adopt systematic uncertainties of 1\% on signal normalization, 0.5\% on energy calibration, and 1\% on background contamination, reflecting JUNO's excellent energy resolution and expected systematic control. 

\begin{figure}[!htb]
    \centering
    \includegraphics[width=\columnwidth]{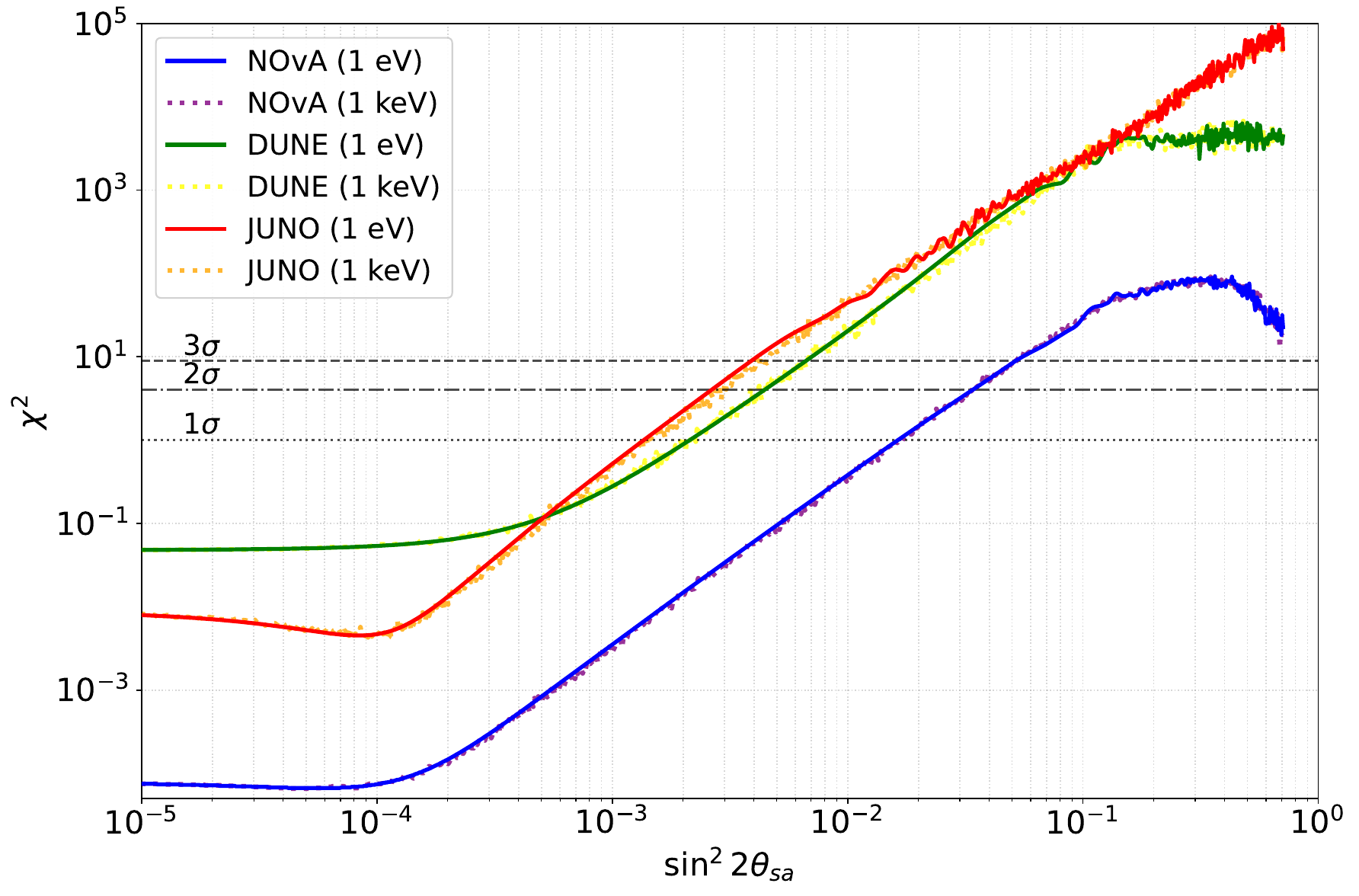}
    \caption{Experimental sensitivity to active--sterile mixing: $\chi^2$ as a function of $\sin^2 2\theta_{sa}$ for NO$\nu$A (dashed), DUNE (dot-dashed), and JUNO (solid), evaluated at the eV (thick lines) and keV (thin lines) sterile mass scales. Horizontal lines mark the $1\sigma$, $2\sigma$, and $3\sigma$ confidence levels ($\Delta\chi^2 = 1, 4, 9$). JUNO provides the strongest exclusion reach, followed by DUNE and NO$\nu$A, reflecting their respective energy resolutions, detector masses, and exposures.}
    \label{fig:chi_square}
\end{figure}

Fig.~(\ref{fig:chi_square}) presents the $\chi^2$ sensitivity as a function of the active–sterile mixing parameter $\sin^2 2\theta_{sa}$ for JUNO, DUNE, and NO$\nu$A, evaluated at both eV and keV sterile neutrino mass scales. Since we vary a single parameter while profiling over systematic uncertainties, the confidence levels correspond to $\Delta\chi^2 = 1$, $4$, and $9$ for $1\sigma$, $2\sigma$, and $3\sigma$ exclusions, respectively.

We observe that JUNO exhibits the strongest sensitivity to sterile neutrino effects, with $\chi^2$ rising steeply even for small values of $\sin^2 2\theta_{sa}$. This behavior is driven by JUNO's high event statistics, exceptional 3\% energy resolution, and the constructive interference between solar and atmospheric oscillation phases at its baseline and energy regime, which amplifies sensitivity to spectral distortions. DUNE demonstrates comparable, though slightly weaker, sensitivity due to its long baseline, large detector mass, and high exposure. NO$\nu$A displays the weakest sensitivity among the three experiments, limited by its smaller detector mass and lower event statistics.

Quantitatively, JUNO and DUNE can exclude active–sterile mixing down to $\sin^2 2\theta_{sa} \sim 10^{-3}$ at the $1\sigma$ level for eV-scale sterile neutrinos, whereas NO$\nu$A becomes sensitive only at larger mixing values of order $10^{-2}$. These results demonstrate the complementary roles of reactor and accelerator-based long-baseline experiments in probing sterile neutrino scenarios across different energy regimes and baselines.

\section{Complementarity with other probes}\label{subsec:complementarity_probes}

\begin{figure*}[!htbp]
    \centering
    \begin{subfigure}[t]{0.48\textwidth}
        \centering
        \includegraphics[width=\linewidth]{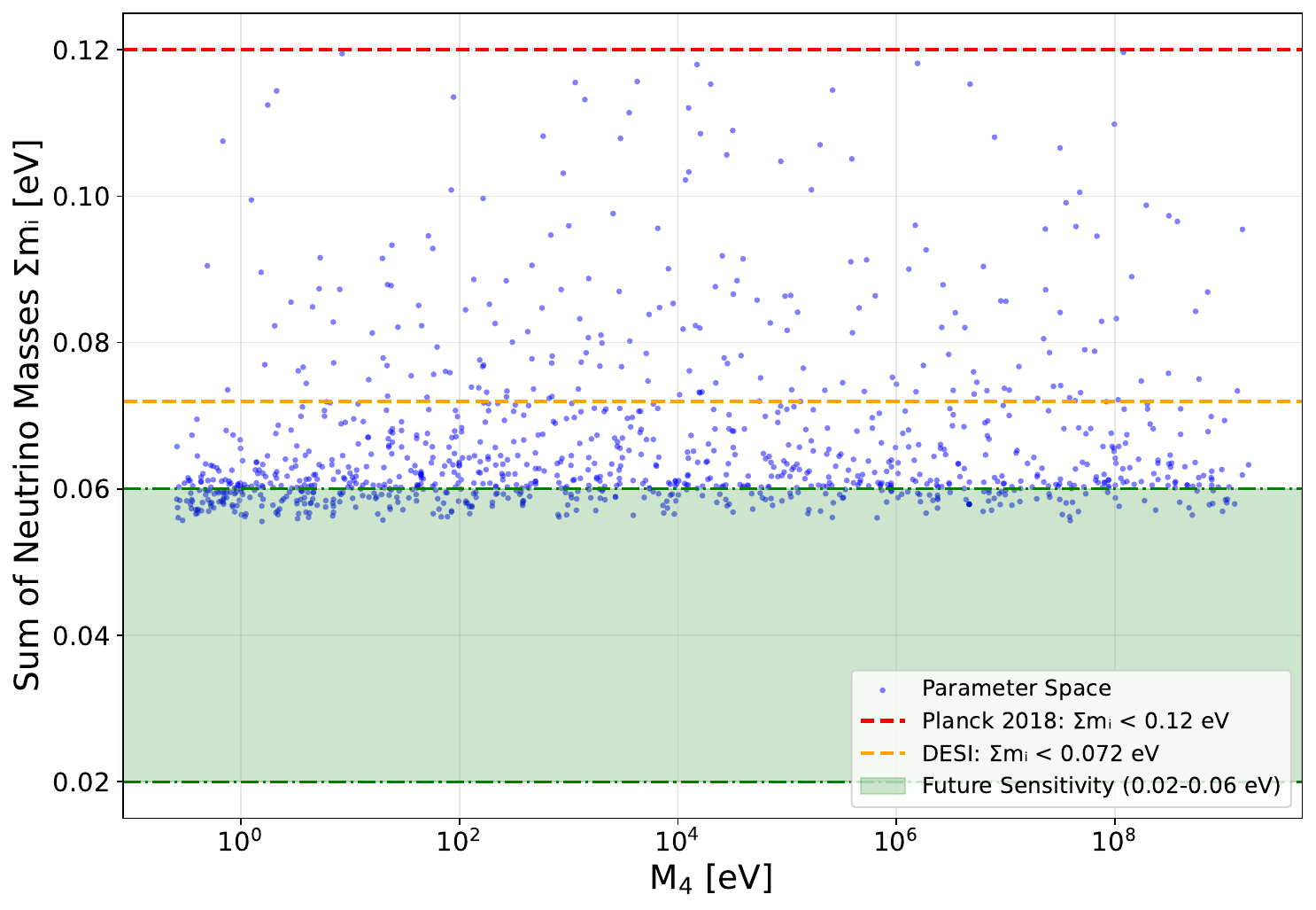}
        \caption{}
        \label{fig:sum_of_nu}
    \end{subfigure}
    \hfill
    \begin{subfigure}[t]{0.48\textwidth}
        \centering
        \includegraphics[width=\linewidth]{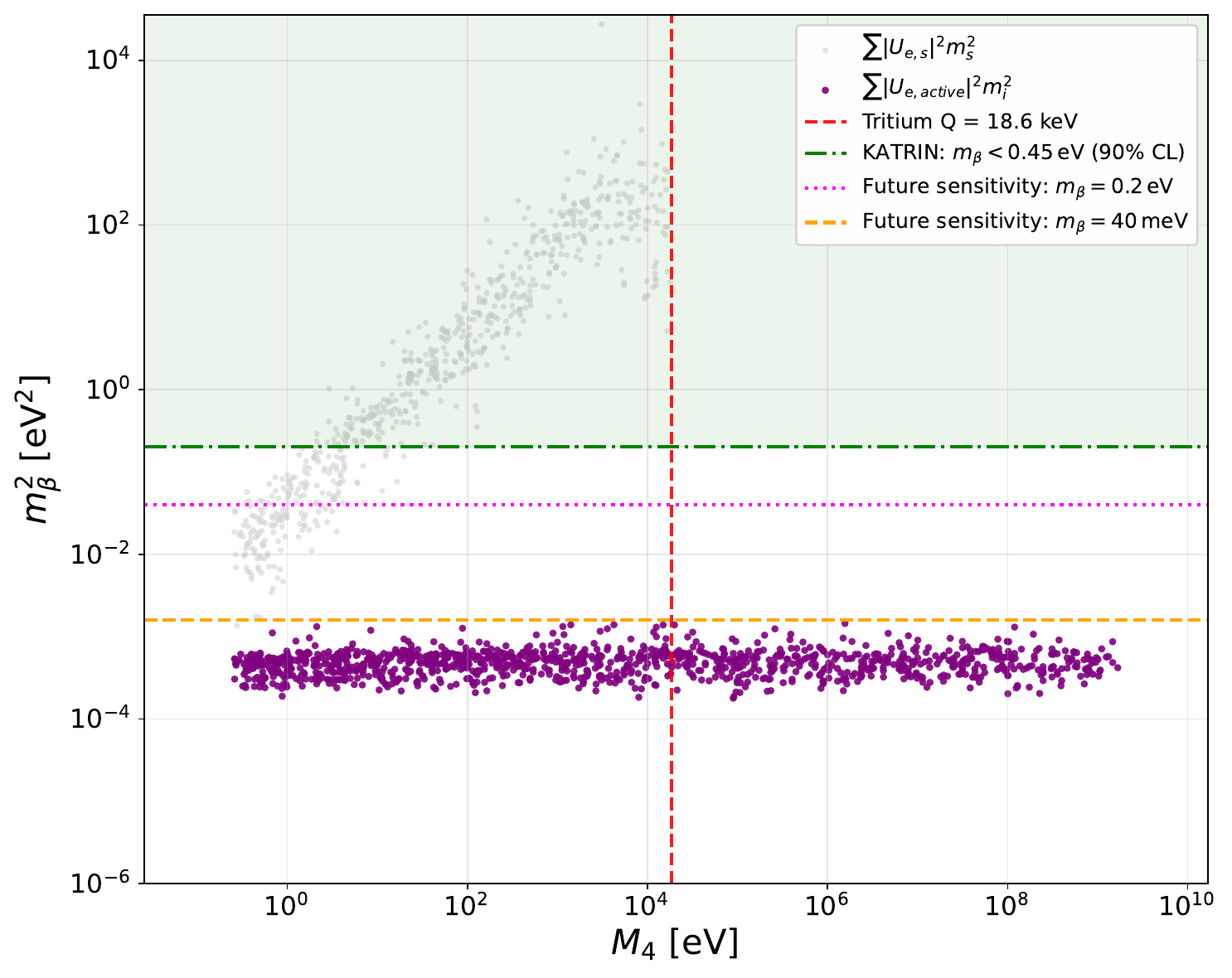}
        \caption{}
        \label{fig:m_beta_values}
    \end{subfigure}
    \caption{Seesaw predictions for the absolute neutrino mass scale. (a)~Predicted sum of neutrino masses $\sum m_i$ as a function of the lightest sterile neutrino mass $M_4$; the horizontal lines indicate the Planck 2018 bound ($<0.12$~eV), the DESI forecast ($<0.072$~eV), and future projections ($<0.06$~eV). (b)~Effective kinematic mass $m_\beta^2$ as a function of $M_4$; the tritium $Q$-value of $18.6$~keV sets a kinematic threshold above which sterile neutrinos cannot contribute to the beta-decay endpoint, with the KATRIN upper limit and future Project~8 sensitivity indicated.}
    \label{fig:mass_probes}
\end{figure*}

\begin{figure*}[!htbp]
    \centering
    \begin{subfigure}[t]{0.48\textwidth}
        \centering
        \includegraphics[width=\linewidth]{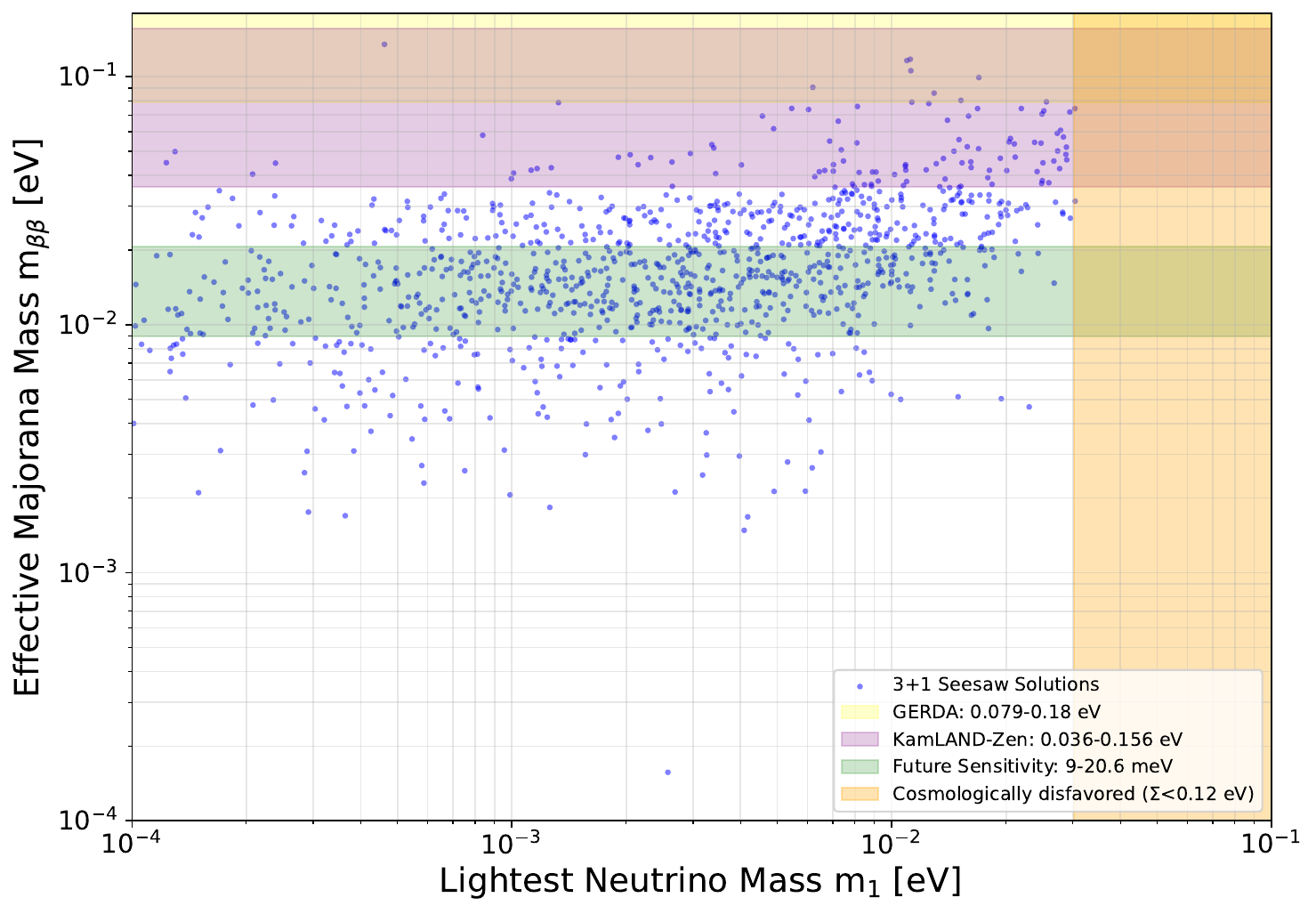}
        \caption{}
        \label{fig:eff_nu_mass}
    \end{subfigure}
    \hfill
    \begin{subfigure}[t]{0.48\textwidth}
        \centering
        \includegraphics[width=\linewidth]{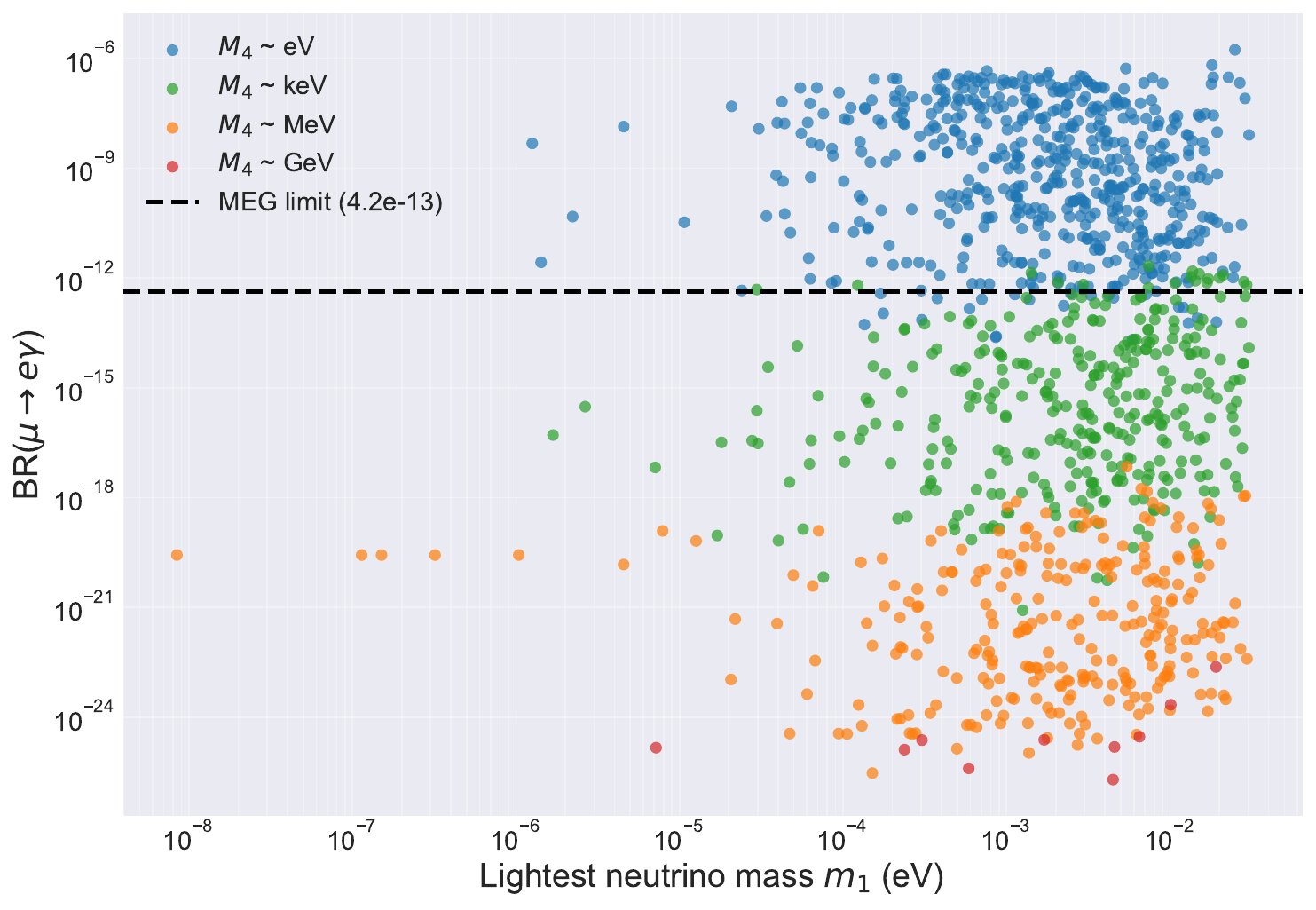}
        \caption{}
        \label{fig:BR_ratio}
    \end{subfigure}
    \caption{Seesaw predictions for lepton-number-violating and flavor-violating observables. (a)~Effective Majorana mass $|m_{\beta\beta}|$ as a function of the lightest neutrino mass $m_1$, with the standard three-flavor bands for normal and inverted hierarchy shown alongside current exclusions from KamLAND-Zen and GERDA and projected sensitivities of LEGEND-1000, nEXO, and CUPID. (b)~Predicted branching ratio $\text{BR}(\mu \to e\gamma)$ as a function of the lightest sterile neutrino mass; the horizontal line marks the current MEG bound ($4.2 \times 10^{-13}$), with eV-scale sterile neutrinos under significant tension.}
    \label{fig:lnv_clfv_probes}
\end{figure*}

While the oscillation analysis of Sec.~(\ref{sec:seesaw_oscillations}) probes the active--sterile mixing through spectral distortions at specific baselines and energies, a complete test of the seesaw mechanism requires confronting its predictions with independent observational channels that are sensitive to different combinations of the neutrino mass and mixing parameters. In this section, we examine four such complementary probes---the cosmological sum of neutrino masses, the kinematic mass from beta decay, the effective Majorana mass from neutrinoless double beta decay ($0\nu\beta\beta$), and the branching ratio of the charged lepton flavor violating (cLFV) process $\mu \to e\gamma$---and compare the seesaw predictions across all four observables simultaneously. The results are collected in Figs.~\ref{fig:mass_probes} and~\ref{fig:lnv_clfv_probes}, which display each observable as a function of the neutrino or sterile neutrino mass, colour-coded by sterile mass scale.

\paragraph{Cosmological sum of neutrino masses.}
Cosmological observations constrain the sum $\sum m_i = m_1 + m_2 + m_3$ through the gravitational effects of relic neutrinos on large-scale structure formation and the CMB~\cite{Planck:2018vyg}. The Planck 2018 analysis establishes an upper bound of $\sum m_i < 0.12$~eV, while forecasts from DESI~\cite{DESI:2016fyo} and upcoming surveys project sensitivities down to $0.06$~eV~\cite{SimonsObservatory:2018koc,Euclid:2024imf,Abazajian:2019eic}. The seesaw predictions, shown in Fig.~\ref{fig:sum_of_nu}, satisfy the Planck bound across the entire scanned parameter space, with the majority of accepted points clustering in the narrow range $0.05$--$0.07$~eV. This concentration arises because the seesaw relation, in conjunction with the oscillation constraints of Table~\ref{tab:constraints}, tightly correlates the sterile-sector parameters with the active mass spectrum. Notably, the predicted range lies squarely within the projected sensitivity of next-generation cosmological surveys, suggesting that a positive measurement of $\sum m_i$ at the level of $\sim 0.06$~eV would be fully consistent with the Type-I seesaw framework.

\paragraph{Kinematic mass from beta decay.}
A model-independent probe of the absolute neutrino mass scale is provided by the kinematic analysis of single beta decay, where the observable~\cite{Xing:2011ur}
\begin{equation}
    m_\beta^2 = \sum_{i=1}^{3}|U_{ei}|^2\, m_i^2
\end{equation}
is extracted from the spectral shape near the endpoint of the tritium decay $^3\text{H} \to {^3\text{He}} + e^- + \bar\nu_e$. The KATRIN experiment~\cite{KATRIN:2024cdt} currently sets $m_\beta < 0.45$~eV at 90\% C.L., with an ultimate sensitivity goal of $\sim 0.2$~eV. As shown in Fig.~\ref{fig:m_beta_values}, the seesaw-predicted $m_\beta^2$ values for the active-only contribution lie well below the current KATRIN sensitivity, and will likely remain inaccessible even to next-generation experiments such as Project~8~\cite{Project8:2022hun}. The situation changes qualitatively, however, when sterile neutrinos are kinematically accessible. For sterile states with masses below the tritium $Q$-value of $18.6$~keV, the beta spectrum acquires a characteristic \textit{kink-like distortion} at an electron energy $E = E_0 - M_j$, arising from the additional incoherent contribution $\sum_{j=4}^{6} |\mathcal{U}_{ej}|^2 M_j^2 \,\Theta(E_0 - M_j)$ to $m_\beta^2$. Crucially, this kink signature is experimentally distinct from the standard endpoint distortion and provides a direct, model-independent avenue for detecting sterile neutrinos through their spectral imprint. KATRIN has already exploited this strategy to search for both eV-scale~\cite{KATRIN:2020dpx} and keV-scale~\cite{KATRIN:2022spi} sterile neutrinos, placing competitive laboratory bounds on the active--sterile mixing amplitude (e.g., $\sin^2\theta < 5 \times 10^{-4}$ at 95\% C.L.\ for sterile masses between 0.1 and 1.0~keV). The upcoming TRISTAN detector upgrade~\cite{KATRIN:2018oow}, designed to measure the full tritium spectrum with a multi-pixel silicon drift detector system, will extend the sensitivity to keV-scale sterile neutrinos by several orders of magnitude. This highlights a key complementarity: while the standard $m_\beta$ measurement probes the active mass spectrum near the endpoint, the kink search across the full beta spectrum provides an independent and uniquely sensitive channel for detecting sterile neutrinos within the seesaw framework.

\paragraph{Effective Majorana mass from $0\nu\beta\beta$ decay.}
If neutrinos are Majorana fermions, the observation of neutrinoless double beta decay would establish lepton number violation and provide a direct measurement of the effective Majorana mass $|m_{\beta\beta}|$~\cite{Das:2023aic,Dolinski:2019nrj}. In the $6\nu$ framework, the effective mass receives contributions from both active and sterile sectors:
\begin{equation}
    |m_{\beta\beta}| = \left|\sum_{i=1}^{3} m_i\,\mathcal{U}_{ei}^2 + \sum_{j=4}^{6} \mathcal{U}_{ej}^2 \frac{M_j\, p^2}{p^2 - M_j^2}\right|,
\end{equation}
where $p^2 \approx -(100~\text{MeV})^2$ is the typical virtual momentum exchanged in the process. For sterile neutrinos lighter than this momentum scale, the sterile contribution adds coherently to the active amplitude; for much heavier sterile states, the propagator suppresses the contribution as $\sim p^2/M_j^2$, consistent with the seesaw decoupling. The predicted values of $|m_{\beta\beta}|$ are presented in Fig.~\ref{fig:eff_nu_mass} as a function of the lightest neutrino mass $m_1$, overlaid with the standard three-flavor normal- and inverted-hierarchy bands. The seesaw predictions populate the same region as the standard bands, but with an additional spread due to the sterile-sector phases. Crucially, many of the predicted points fall within the projected sensitivity of next-generation experiments such as LEGEND-1000, nEXO, and CUPID, making $0\nu\beta\beta$ searches a powerful discriminator of the seesaw parameter space.

\paragraph{Charged lepton flavor violation: $\text{BR}(\mu \to e\gamma)$.}
In the SM with Dirac neutrino masses, the cLFV decay $\mu \to e\gamma$ is suppressed to unobservable levels ($\sim 10^{-54}$) by the GIM mechanism~\cite{Casas:2001sr}. The seesaw mechanism lifts this suppression dramatically, as the heavy sterile neutrinos propagate in the loop mediating $\ell \to \ell'\gamma$ with a coupling proportional to the active--sterile mixing~\cite{Ilakovac:1994kj,Abada:2015oba}. Non-unitarity of the PMNS matrix provides a convenient parametrization of these effects. For low-scale seesaw scenarios with heavy neutrinos at or below the electroweak scale, the branching ratio is given by~\cite{Garnica:2023ccx}
\begin{align}
    \text{BR}(\ell \to \ell' \gamma) &= \frac{\alpha}{\Gamma_\ell}\, m_\ell^3\, |F_{\gamma M}^\ell(0)|^2, \\
    F_{\gamma M}^\ell(0) &= \frac{\alpha_W}{16\pi} \frac{m_\ell}{M_W^2} \sum_i B_{\ell i}^* B_{\ell' i}\, f_{\gamma M}^\ell(x_i), \\
    f_{\gamma M}^\ell(x) &= \frac{3x^3 \log x}{2(x-1)^4} - \frac{2x^3 + 5x^2 - x}{4(x-1)^3} + \frac{5}{6}, \nonumber
\end{align}
where $\alpha_W = \alpha/\sin^2\theta_W$, $x_i = m_{\chi_i}^2/M_W^2$, and $m_{\chi_i}$ denotes the mass of each physical neutrino state. The current MEG bound $\text{BR}(\mu \to e\gamma) < 4.2 \times 10^{-13}$~\cite{MEGII:2023ltw} already provides meaningful constraints on the active--sterile mixing. The predicted branching ratios, shown in Fig.~\ref{fig:BR_ratio}, reveal a striking mass-scale dependence. For sterile neutrinos in the keV range and above, the predicted $\text{BR}(\mu \to e\gamma)$ values lie comfortably below the MEG bound, consistent with the expected decoupling behavior of the loop function. In contrast, for eV-scale sterile neutrinos, the relatively larger active--sterile mixing angles required by the seesaw relation push the predicted branching ratios to values that approach or exceed the current experimental bound. This implies that eV-scale sterile neutrinos within the seesaw framework are already under significant tension from cLFV constraints, and that the effects of such light sterile states would be among the most readily observable signatures of the seesaw mechanism.
\section{Some model-building aspects of Low-Scale Seesaw Models}
\label{sec:low_scale_justification}
A significant theoretical challenge for low-scale seesaw models is the "naturalness" problem. The standard seesaw relation ($m_\nu \approx m_D^2/M_R$) implies that if the right-handed neutrino mass scale ($M_R$) is low (e.g., eV to TeV scale), the Dirac mass term ($m_D$) must be exceptionally small to reproduce the observed light neutrino masses. This corresponds to tiny Yukawa couplings ($Y_\nu = m_D/v$, where $v$ is the Higgs VEV), which are often considered unnatural. However, several well-motivated theoretical frameworks have been proposed to address this issue and provide a robust justification for low-scale seesaw scenarios. These models typically introduce new physics at the TeV scale, making them potentially testable at colliders.

One prominent approach involves extending the gauge group of the Standard Model. For instance, by introducing a local $U(1)_{B-L}$ (Baryon-Lepton number) gauge symmetry that is broken at a scale far below the GUT scale, it is possible to generate Majorana masses for right-handed neutrinos naturally, without requiring a fundamental high-energy scale~\cite{Debnath:2023akj}.
Another class of models introduces new particle content. For example, in 3-3-1 models (based on an $SU(3)_C \times SU(3)_L \times U(1)_X$ gauge group), the introduction of a scalar sextet can naturally generate right-handed neutrino masses in the keV-GeV range, providing a consistent framework for sterile neutrino phenomenology~\cite{Cabrera:2023rcy}.
Other mechanisms rely on specific symmetry-breaking patterns. Spontaneous breaking of a discrete symmetry at an energy scale much lower than the fundamental cutoff can give rise to right-handed neutrino masses around the 50 GeV scale, making them accessible to collider searches~\cite{CarcamoHernandez:2019kjy}. Supersymmetric models also offer solutions where the B-L gauge symmetry breaking scale is disentangled from the right-handed neutrino mass scale, allowing for eV-scale neutrino masses~\cite{Dib:2019jod,Borzumati:2000mc}.

These models not only provide a theoretical foundation for low-scale seesaw mechanisms but also connect sterile neutrino physics to other important areas of particle physics and cosmology, such as baryogenesis. For example, models with right-handed neutrinos below the pion mass have been studied in the context of their decay to Standard Model particles, exploring whether they can account for the observed Baryon Asymmetry of the Universe (BAU)~\cite{Domcke:2020ety}.

Collectively, these approaches demonstrate that low-scale seesaw models can be theoretically well-motivated, testable, and connected to a broader landscape of BSM physics. 
\section{Conclusion}\label{sec:conclusion}
The central question addressed in this work was to what extent the high-scale structure of the Type-I seesaw mechanism can be tested indirectly through low-energy observables. While the sterile sector is not directly accessible at current collider energies, its parameters leave calculable imprints on the light-neutrino masses, mixing angles, CP-violating phases, and deviations from unitarity. By constructing an explicit bridge between the sterile-sector parameters and the effective oscillation observables through the exact seesaw relation, we have systematically explored how much of the underlying parameter space can be constrained by current and near-future experimental data.

Our analysis reveals that the allowed parameter space, far from being arbitrary, is highly structured. The characteristic inverse scaling between sterile mass and active--sterile mixing governs the viable regions and naturally leads to the decoupling of very heavy sterile states from oscillation phenomenology. Long- and medium-baseline experiments---DUNE, NO$\nu$A, and JUNO---are primarily sensitive to the lower-mass sterile regime, where mixing effects remain dynamically relevant and produce measurable distortions in oscillation probabilities and event rates. In particular, JUNO exhibits the strongest sensitivity owing to its exceptional energy resolution and high statistics, followed by DUNE with its long baseline and large exposure. At larger sterile mass scales, the oscillatory signatures average out, and the sterile states contribute only through suppressed non-unitarity corrections, making their direct oscillation signatures effectively unobservable but leaving imprints in other channels.

The complementary probes examined in Sec.~(\ref{subsec:complementarity_probes}) substantially reinforce and extend the oscillation analysis. Taken together, the four observables---the cosmological sum of neutrino masses, the kinematic mass $m_\beta$ from beta decay, the effective Majorana mass $|m_{\beta\beta}|$ from $0\nu\beta\beta$ decay, and the cLFV branching ratio $\text{BR}(\mu \to e\gamma)$---paint a consistent picture of the seesaw parameter space. The cosmological and $m_\beta$ observables primarily constrain the \textit{active} mass spectrum: the seesaw predictions cluster tightly at $\sum m_i \sim 0.05$--$0.07$~eV, squarely within the sensitivity of next-generation cosmological surveys, while the standard $m_\beta$ values remain below current KATRIN reach. However, the kink signatures from kinematically accessible sterile states provide a distinct detection channel that KATRIN has already begun to exploit, with the upcoming TRISTAN upgrade projected to extend this sensitivity dramatically. The $0\nu\beta\beta$ effective mass bridges both sectors, with many seesaw-predicted $|m_{\beta\beta}|$ values falling within the projected sensitivities of LEGEND-1000, nEXO, and CUPID. The cLFV branching ratio provides the most direct probe of the active--sterile mixing magnitude: for eV-scale sterile neutrinos, the predicted $\text{BR}(\mu \to e\gamma)$ values approach or exceed the current MEG bound, placing this mass regime under significant tension, whereas keV-scale and heavier sterile states decouple progressively from the cLFV constraints. The interplay of all these probes significantly reduces the otherwise vast seesaw parameter space, though important viable regions persist---highlighting both the flexibility of the seesaw framework and the intrinsic limitations of any single experimental channel in fully resolving the heavy sector.

Several extensions of the present analysis remain open. We have assumed stable sterile states and neglected possible decay channels, which could introduce damping effects and modified phase evolution in oscillation probabilities. The implications for leptogenesis, direct collider signatures, and astrophysical neutrino constraints have not been addressed, each of which may further reshape the allowed parameter space. A fully dynamical treatment incorporating lifetime effects, thermal history considerations, and the complete set of non-oscillation observables would provide a more comprehensive assessment of the phenomenological viability of the seesaw mechanism.

In summary, while oscillation experiments alone cannot definitively determine the high-scale seesaw parameters, they impose nontrivial and structured constraints on the sterile sector that, when combined with cosmological, $0\nu\beta\beta$, kinematic, and cLFV data, significantly narrow the allowed parameter space. The framework developed here provides a consistent basis for translating future experimental precision into progressively sharper tests of neutrino mass generation mechanisms. As next-generation facilities---DUNE, JUNO, LEGEND-1000, Project~8, and MEG~II---improve sensitivity to mixing parameters and absolute mass scales, the indirect window into the seesaw structure will continue to narrow, bringing us closer to a definitive test of the origin of neutrino mass.

\newpage
\appendix
\textbf{\Huge Appendix}
\section{Seesaw Relation}\label{sec:seesaw_relation_appendix}
The introduction of the right-handed neutrinos into the Lagrangian that couples with the Higgs particle to generate mass results in the formation of the mass matrix given as Eq.~(\ref{eq:mass_Matrix}). This mass is the effective mass of the interacting state — that is, the flavour state. The mixing between neutrinos is defined as, 
\begin{equation}
    \nu_f = \mathcal{U}\times\nu_m
\end{equation}
So, when we change to the mass basis, we get the mass Lagrangian of the form,
\begin{equation}
    \mathcal{L}_{mass} = \psi_L^TC^{-1}\mathcal{M}\psi_L+h.c. = (\mathcal{U}\chi_L)^TC^{-1}\mathcal{M}(\mathcal{U}\chi_L)
\end{equation} 
where $\psi_L = \begin{pmatrix}
    \nu_L&\nu_R^C
\end{pmatrix}^T$ and $\chi_L = \begin{pmatrix}
    \nu_{mL}&\nu_{mR}
\end{pmatrix}^T$. This leads to the effective mass matrix being $\mathcal{U}^T\mathcal{MU}$. Since $\mathcal{M}$ is a complex-symmetric matrix, by Takagi's factorization we get, 
\begin{equation}
    \mathcal{U}^T\mathcal{MU} = \mathcal{D}
\end{equation}
\begin{widetext}
\begin{align}
    \implies\begin{pmatrix}
        N^T&T^T\\R^T&B^T
    \end{pmatrix}\begin{pmatrix}
        0&m_D\\m_D^T&M_R
    \end{pmatrix}\begin{pmatrix}
        N&R\\T&B
    \end{pmatrix} = \begin{pmatrix}
        \mathcal{D_\nu}&0\\0&\mathcal{D}_N
    \end{pmatrix}
    \label{eq:UMUD}
\end{align}
With some mathematical manipulation, we get,
\begin{equation}
\begin{pmatrix}
    0&m_D\\m_D^T&M_R
\end{pmatrix}=
    \begin{pmatrix}
        N^*&R^*\\T^*&B^*
    \end{pmatrix}\begin{pmatrix}
        \mathcal{D_\nu}&0\\0&\mathcal{D}_N
    \end{pmatrix}\begin{pmatrix}
        N^\dagger&T^\dagger\\R^\dagger&B^\dagger
    \end{pmatrix}\label{eq:seesaw_matrix}
\end{equation}
\end{widetext}
Comparing the block matrices of LHS and RHS, we get, 
\begin{align}
    N^*\mathcal{D_\nu}N^\dagger + R^*\mathcal{D}_NR^\dagger = 0\\
    \implies N\mathcal{D_\nu}N^T + R\mathcal{D}_NR^T = 0\\
    \implies N\mathcal{D_\nu}N^T = - R\mathcal{D}_NR^T
\end{align}
Using $N = AU_0$ in the equation, we arrive at the final form of the seesaw relation as, 
\begin{equation}
    U_0\mathcal{D_\nu}U_0^T = - (A^{-1}R)\mathcal{D}_N{(A^{-1}R)}^T
\end{equation}

\section{Mass Matrix Diagonalisation}\label{sec:mass_matrix_diagonalisation}
The mass matrix $\mathcal{M}$ is a block diagonal matrix. So it can be diagonalized by using the Schur complement. Since $M_R\neq0$ and is invertible, then by the Schur complement method, the diagonalized matrix can be calculated as, 
\begin{widetext}
\begin{align}
    \begin{pmatrix}
        \mathcal{D_\nu}&0\\0&\mathcal{D}_N
    \end{pmatrix} = \begin{pmatrix}
        \mathcal{I}& -m_D M_R^{-1}\\0&\mathcal{I}
    \end{pmatrix}\begin{pmatrix}
        0&m_D\\m_D^T&M_R
    \end{pmatrix}\begin{pmatrix}
        \mathcal{I}&0\\-M_R^{-1}m_D^T&\mathcal{I}
    \end{pmatrix}
\end{align}
\end{widetext}
This leads to the form of the neutrino mass in terms of Dirac mass and Majorana mass becoming, 
\begin{equation}
    \begin{pmatrix}
        \mathcal{D_\nu}&0\\0&\mathcal{D}_N
    \end{pmatrix} = \begin{pmatrix}
        -m_DM_R^{-1}m_D^T&0\\0&M_R
    \end{pmatrix}
\end{equation}

\section{Neutrino Mass and Standard Neutrino Mixing from Seesaw Relation}\label{sec:nu_mass_mix_appendix}
Using the Hermitian matrix from Eq.~(\ref{eq:herm_mat}), we can analytically calculate the mass of the light neutrinos as a function of sterile neutrino parameters. For that, we take the Hermitian matrix and find the eigenvalues.

For a 3$\times$3 matrix, the eigenvalue equation corresponds to that of a cubic equation. So for eigenvalue $\lambda$, we will have the equation as, 
\begin{equation}
\lambda^3 - Tr(H)\lambda^2+\frac{[Tr(H)]^2-Tr(H^2)}{2}\lambda - Det(H) = 0
\label{eq:cubic_eigen}
\end{equation}
Since the eigenvalues are mass-squared values, that makes $\lambda = m_i^2$. Taking the Hermitian matrix from the RHS of Eq.~(\ref{eq:herm_mat}), the mass-squared values can be calculated from the sterile neutrino parameters.

The roots of Eq.~(\ref{eq:cubic_eigen}) give, 
\begin{widetext}
\begin{equation}
\begin{aligned}
m_1^{2} &= 
\frac{1}{3}\operatorname{Tr}(H)
-\frac{1}{3}
\sqrt{\tfrac12\big(3\operatorname{Tr}(H^{2})-[\operatorname{Tr}(H)]^{2}\big)}
\left( \cos\Phi + \sqrt{3}\,\sin\Phi \right),
\end{aligned}
\end{equation}

\begin{equation}
\begin{aligned}
m_2^{2} &= 
\frac{1}{3}\operatorname{Tr}(H)
-\frac{1}{3}
\sqrt{\tfrac12\big(3\operatorname{Tr}(H^{2})-[\operatorname{Tr}(H)]^{2}\big)}
\left( \cos\Phi - \sqrt{3}\,\sin\Phi \right),
\end{aligned}
\end{equation}

\begin{equation}
\begin{aligned}
m_3^{2} &= 
\frac{1}{3}\operatorname{Tr}(H)
+\frac{2}{3}\cos\Phi\,
\sqrt{\tfrac12\big(3\operatorname{Tr}(H^{2})-[\operatorname{Tr}(H)]^{2}\big)}.
\end{aligned}
\end{equation}
\end{widetext}

From the unitarity relations and the exact seesaw relations, we have a set of equations to solve for each PMNS matrix element based on the sterile parameters as, 
\begin{widetext}
\begin{align}
(U_0)_{\alpha 1} (U_0)_{\beta 1}^* + (U_0)_{\alpha 2} (U_0)_{\beta 2}^* + (U_0)_{\alpha 3} (U_0)_{\beta 3}^* &= \delta_{\alpha\beta} \nonumber \\
m_1^2 (U_0)_{\alpha 1} (U_0)_{\beta 1}^* + m_2^2 (U_0)_{\alpha 2} (U_0)_{\beta 2}^* + m_3^2 (U_0)_{\alpha 3} (U_0)_{\beta 3}^* &= H_{\alpha\beta} \nonumber \\
m_1^4 (U_0)_{\alpha 1} (U_0)_{\beta 1}^* + m_2^4 (U_0)_{\alpha 2} (U_0)_{\beta 2}^* + m_3^4 (U_0)_{\alpha 3} (U_0)_{\beta 3}^* &= (H^2)_{\alpha\beta} \label{eq:unitary_relations}
\end{align}

Solving the three equations, we can get the expression of the mixing matrix elements as, 
\begin{align}
    |(U_0)_{\alpha 1}|^2 = \frac{(H^2)_{\alpha\alpha}-H_{\alpha\alpha}(m_2^2+m_3^2)+m_2^2m_3^2}{\Delta m^2_{21}\Delta m^2_{31}}\\
    |(U_0)_{\alpha 2}|^2 = \frac{(H^2)_{\alpha\alpha}-H_{\alpha\alpha}(m_1^2+m_3^2)+m_1^2m_3^2}{\Delta m^2_{21}\Delta m^2_{23}}\\
    |(U_0)_{\alpha 3}|^2 = \frac{(H^2)_{\alpha\alpha}-H_{\alpha\alpha}(m_1^2+m_2^2)+m_1^2m_2^2}{\Delta m^2_{31}\Delta m^2_{32}}\label{eq:U_H_relation}
\end{align}
\end{widetext}

Using these solutions, we can calculate the mixing angles as, 
\begin{align}
    \theta_{12} &= \sin^{-1}\sqrt{\left(\frac{|(U_0)_{e2}|^2}{1-|(U_0)_{e3}|^2}\right)}, \nonumber\\
    \theta_{13} &= \sin^{-1}|(U_0)_{e3}|, \nonumber\\
    \theta_{23} &= \sin^{-1}\sqrt{\left(\frac{|(U_0)_{\mu3}|^2}{1-|(U_0)_{e3}|^2}\right)}.
\end{align}

Using the derived relations from Eq.~(\ref{eq:U_H_relation}), we can obtain the exact value and form of the mixing angles based on the sterile neutrino parameters. The phases of the active neutrinos can be calculated similarly as, 
\begin{align}
    \delta_{12} &= -\tan^{-1}\left(\frac{\Im(U_0)_{12}}{\Re(U_0)_{12}}\right), \nonumber\\
    \delta_{23} &= -\tan^{-1}\left(\frac{\Im(U_0)_{23}}{\Re(U_0)_{23}}\right), \nonumber\\
    \delta_{13} &= \tan^{-1}\left(\frac{\Im(U_0)_{13}}{\Re(U_0)_{13}}\right).
\end{align}

\section{Mixing Matrices}
\label{sec:Mixing Matrices}
The $6\times6$ mixing matrix is formed as mentioned in Eq.~(\ref{eq:Unitary_matrix_representation}). Here we show the explicit forms of the block mixing matrices that we introduced in Sec.~(\ref{sec:seesaw_mechanism}). 
The active-active mixing is defined as $U_0$ = $O_{23}O_{13}O_{12}$. This is the PMNS matrix and it's form is given as 
\begin{widetext}
\begin{equation}
   U_0 = \begin{pmatrix}
c_{12} c_{13} &
c_{13} \hat{s}_{12} &
\hat{s}_{13} \\

- c_{23} \hat{s}_{12}^{*}
- c_{12} \hat{s}_{13} \hat{s}_{23}^{*} &
c_{12} c_{23}
- \hat{s}_{12} \hat{s}_{13} \hat{s}_{23}^{*} &
c_{13} \hat{s}_{23} \\

\hat{s}_{12}^{*} \hat{s}_{23}
- c_{12} c_{23} \hat{s}_{13}^{*} &
- c_{12} \hat{s}_{23}
- c_{23} \hat{s}_{12} \hat{s}_{13}^{*} &
c_{13} c_{23}
\end{pmatrix}
\end{equation}
where $c_{ij} = \cos{\theta_{ij}}$ and $\hat{s}_{ij}=\sin{\theta_{ij}e^{i\delta_{ij}}}$. 
\end{widetext}
The sterile-sterile mixing box is similarly defined as $U_{ss}$ = $O_{56}O_{46}O_{45}$. This gives the sterile-sterile mixing block as, 
\begin{widetext}
\begin{equation}
U_{ss} = \begin{pmatrix}
c_{45} c_{46}
&
c_{46} \hat{s}_{45}
&
\hat{s}_{46}
\\

- c_{56} \hat{s}_{45}^{*}
- c_{45} \hat{s}_{46} \hat{s}_{56}^{*}
&
c_{45} c_{56}
- \hat{s}_{45} \hat{s}_{46} \hat{s}_{56}^{*}
&
c_{46} \hat{s}_{56}
\\

\hat{s}_{45}^{*} \hat{s}_{56}
- c_{45} c_{56} \hat{s}_{46}^{*}
&
- c_{45} \hat{s}_{56}
- c_{56} \hat{s}_{45} \hat{s}_{46}^{*}
&
c_{46} c_{56}
\end{pmatrix}
\end{equation}
\end{widetext}
There are 4 active-sterile mixing blocks. These mixings are also the one that induces the non-unitarity in the active-active and sterile-sterile mixing blocks. The explicit expression of these active-sterile mixing blocks is given as,
\begin{widetext}
\begin{equation}
    A = \begin{pmatrix}
c_{14} c_{15} c_{16} & 0 & 0 \\

\begin{aligned}
&-\hat{s}^*_{14} \hat{s}_{24} c_{25} c_{26} \\
&- c_{14} ( \hat{s}^*_{15} \hat{s}_{25} c_{26} + \hat{s}^*_{16} \hat{s}_{26} c_{15} )
\end{aligned} & 
c_{24} c_{25} c_{26} & 
0 \\

\begin{aligned}
&-\hat{s}^*_{14} [ \hat{s}_{34} c_{24} c_{35} c_{36} \\
&\quad - \hat{s}_{24} (\hat{s}^*_{25} \hat{s}_{35} c_{36} + \hat{s}^*_{26} \hat{s}_{36} c_{25}) ] \\
&- c_{14} [ \hat{s}^*_{16} \hat{s}_{36} c_{15} c_{26} \\
&\quad + \hat{s}^*_{15} (\hat{s}_{35} c_{25} c_{36} - \hat{s}_{25} \hat{s}^*_{26} \hat{s}_{36}) ]
\end{aligned} & 
\begin{aligned}
&-\hat{s}^*_{24} \hat{s}_{34} c_{35} c_{36} \\
&- c_{24} ( \hat{s}^*_{25} \hat{s}_{35} c_{36} + \hat{s}^*_{26} \hat{s}_{36} c_{25} )
\end{aligned} & 
c_{34} c_{35} c_{36}
\end{pmatrix}
\end{equation}

\begin{equation}
    R = \begin{pmatrix}
\hat{s}_{14} c_{15} c_{16} & \hat{s}_{15} c_{16} & \hat{s}_{16} \\

\begin{aligned}
&\hat{s}_{24} c_{14} c_{25} c_{26} \\
&- \hat{s}_{14} ( \hat{s}^*_{15} \hat{s}_{25} c_{26} + \hat{s}^*_{16} \hat{s}_{26} c_{15} )
\end{aligned} & 
\begin{aligned}
&\hat{s}_{25} c_{15} c_{26} \\
&- \hat{s}_{15} \hat{s}^*_{16} \hat{s}_{26}
\end{aligned} & 
\hat{s}_{26} c_{16} \\

\begin{aligned}
&c_{14} [ \hat{s}_{34} c_{24} c_{35} c_{36} \\
&\quad - \hat{s}_{24} (\hat{s}^*_{25} \hat{s}_{35} c_{36} + \hat{s}^*_{26} \hat{s}_{36} c_{25}) ] \\
&- \hat{s}_{14} [ \hat{s}^*_{16} \hat{s}_{36} c_{15} c_{26} \\
&\quad + \hat{s}^*_{15} (\hat{s}_{35} c_{25} c_{36} - \hat{s}_{25} \hat{s}^*_{26} \hat{s}_{36}) ]
\end{aligned} & 
\begin{aligned}
&-\hat{s}_{15} \hat{s}^*_{16} \hat{s}_{36} c_{26} \\
&+ c_{15} ( \hat{s}_{35} c_{25} c_{36} - \hat{s}_{25} \hat{s}^*_{26} \hat{s}_{36} )
\end{aligned} & 
\hat{s}_{36} c_{16} c_{26}
\end{pmatrix}
\end{equation}

\begin{equation}
    S = \begin{pmatrix}
-\hat{s}^*_{14} c_{24} c_{34} & -\hat{s}^*_{24} c_{34} & -\hat{s}^*_{34} \\

\begin{aligned}
&-\hat{s}^*_{15} c_{14} c_{25} c_{35} \\
&- \hat{s}^*_{14} ( -\hat{s}_{24} \hat{s}^*_{25} c_{35} - \hat{s}_{34} \hat{s}^*_{35} c_{24} )
\end{aligned} & 
\begin{aligned}
&-\hat{s}^*_{25} c_{24} c_{35} \\
&+ \hat{s}^*_{24} \hat{s}_{34} \hat{s}^*_{35}
\end{aligned} & 
-\hat{s}^*_{35} c_{34} \\

\begin{aligned}
&c_{14} [ -\hat{s}^*_{16} c_{15} c_{26} c_{36} \\
&\quad - \hat{s}^*_{15} ( -\hat{s}_{25} \hat{s}^*_{26} c_{36} - \hat{s}_{35} \hat{s}^*_{36} c_{25} ) ] \\
&- \hat{s}^*_{14} [ -\hat{s}_{34} \hat{s}^*_{36} c_{24} c_{35} \\
&\quad + \hat{s}_{24} ( -\hat{s}^*_{26} c_{25} c_{36} + \hat{s}^*_{25} \hat{s}_{35} \hat{s}^*_{36} ) ]
\end{aligned} & 
\begin{aligned}
&\hat{s}^*_{24} \hat{s}_{34} \hat{s}^*_{36} c_{35} \\
&+ c_{24} ( -\hat{s}^*_{26} c_{25} c_{36} + \hat{s}^*_{25} \hat{s}_{35} \hat{s}^*_{36} )
\end{aligned} & 
-\hat{s}^*_{36} c_{34} c_{35}
\end{pmatrix}
\end{equation}

\begin{equation}
    B = \begin{pmatrix}
c_{14} c_{24} c_{34} & 0 & 0 \\

\begin{aligned}
&-\hat{s}_{14} \hat{s}^*_{15} c_{25} c_{35} \\
&+ c_{14} ( -\hat{s}_{24} \hat{s}^*_{25} c_{35} - \hat{s}_{34} \hat{s}^*_{35} c_{24} )
\end{aligned} & 
c_{15} c_{25} c_{35} & 
0 \\

\begin{aligned}
&\hat{s}_{14} [ -\hat{s}^*_{16} c_{15} c_{26} c_{36} \\
&\quad - \hat{s}^*_{15} ( -\hat{s}_{25} \hat{s}^*_{26} c_{36} - \hat{s}_{35} \hat{s}^*_{36} c_{25} ) ] \\
&+ c_{14} [ -\hat{s}_{34} \hat{s}^*_{36} c_{24} c_{35} \\
&\quad + \hat{s}_{24} ( -\hat{s}^*_{26} c_{25} c_{36} + \hat{s}^*_{25} \hat{s}_{35} \hat{s}^*_{36} ) ]
\end{aligned} & 
\begin{aligned}
&-\hat{s}_{15} \hat{s}^*_{16} c_{26} c_{36} \\
&+ c_{15} ( -\hat{s}_{25} \hat{s}^*_{26} c_{36} - \hat{s}_{35} \hat{s}^*_{36} c_{25} )
\end{aligned} & 
c_{16} c_{26} c_{36}
\end{pmatrix}
\end{equation}

The matrices $ A$ and $B$ are lower triangular matrices. These are the matrices that lead to the non-unitary being induced in the active-active mixing ($U_0$) and sterile-sterile mixing ($U_{ss}$) blocks, respectively. 
\end{widetext}
\section{Seesaw Parameter Space}\label{sec:seesaw_param_space}
Table~\ref{tab:sterile_samples} shows one set of sterile parameter space and their corresponding resulting standard neutrino oscillation values lying within $3\sigma$ bounds of NuFIT 6.0 \cite{Esteban:2024eli} for different mass scales of sterile neutrinos. These parameter spaces were obtained through Monte-Carlo simulations of the whole sterile parameter space that will result in the standard neutrino parameters obeying the constraints of Table~\ref{tab:constraints} within $3\sigma$. 
\begin{sidewaystable*}[p]
\centering
\caption{Seesaw parameter sets derived from Monte Carlo Simulations across four sterile mass scales and the resulting derived standard oscillation parameters. The three rows within each block correspond to sterile indices $i=4,5,6$. Masses $M_i$ are in eV, mixing angles $\theta_{ji}$ and phases $\delta_{ji}$
are in radians, and derived angles and CP phase in degrees.}
\label{tab:sterile_samples}
\renewcommand{\arraystretch}{1.4}
\setlength{\tabcolsep}{5pt}
\resizebox{\textheight}{!}{%
\begin{tabular}{|c|r|c|c|c|c|c|c||c|c|c|c|c|c|}
\hline
\multirow{2.5}{*}{\textbf{Scale}} &
\multicolumn{7}{c||}{\textbf{Seesaw parameters}} &
\multicolumn{6}{c|}{\textbf{Derived standard oscillation parameters}} \\
\cline{2-14}
& \multicolumn{1}{c|}{$M_i$\,[eV]}
& $\theta_{1i}$ & $\theta_{2i}$ & $\theta_{3i}$
& $\delta_{1i}$ & $\delta_{2i}$ & \multicolumn{1}{c||}{$\delta_{3i}$}
& $\dfrac{\Delta m^2_{21}}{10^{-5}\,\text{eV}^2}$
& $\dfrac{\Delta m^2_{31}}{10^{-3}\,\text{eV}^2}$
& $\theta_{12}[^\circ]$ & $\theta_{13}[^\circ]$
& $\theta_{23}[^\circ]$ & $\delta_{\mathrm{CP}}[^\circ]$ \\
\hline\hline
\multirow{3}{*}{eV}
 & $9.93\times10^{-1}$ & $2.99\times10^{-2}$ & $1.10\times10^{-1}$ & $1.12\times10^{-1}$ & $1.37$ & $5.61$ & $2.51$
 & \multirow{3}{*}{$7.53$} & \multirow{3}{*}{$2.48$}
 & \multirow{3}{*}{$34.7$} & \multirow{3}{*}{$8.79$} & \multirow{3}{*}{$45.3$} & \multirow{3}{*}{$114$} \\
\cline{2-8}
 & $2.38$ & $9.02\times10^{-3}$ & $9.78\times10^{-2}$ & $1.29\times10^{-1}$ & $0.917$ & $1.58$ & $0.820$ &&&&&&\\
\cline{2-8}
 & $5.51$ & $2.97\times10^{-2}$ & $3.47\times10^{-2}$ & $1.97\times10^{-2}$ & $4.55$ & $4.20$ & $2.19$ &&&&&&\\
\hline
\multirow{3}{*}{keV}
 & $3.10\times10^{3}$ & $3.25\times10^{-4}$ & $2.29\times10^{-3}$ & $5.46\times10^{-4}$ & $6.28$ & $2.40$ & $5.07$
 & \multirow{3}{*}{$7.20$} & \multirow{3}{*}{$2.44$}
 & \multirow{3}{*}{$35.5$} & \multirow{3}{*}{$8.96$} & \multirow{3}{*}{$41.4$} & \multirow{3}{*}{$336$} \\
\cline{2-8}
 & $5.91\times10^{3}$ & $2.08\times10^{-4}$ & $2.54\times10^{-3}$ & $2.69\times10^{-3}$ & $3.84$ & $5.17$ & $3.02$ &&&&&&\\
\cline{2-8}
 & $1.25\times10^{4}$ & $6.24\times10^{-4}$ & $1.48\times10^{-3}$ & $9.49\times10^{-4}$ & $5.78$ & $0.650$ & $1.45$ &&&&&&\\
\hline
\multirow{3}{*}{MeV}
 & $3.10\times10^{6}$ & $3.11\times10^{-5}$ & $1.21\times10^{-5}$ & $8.41\times10^{-5}$ & $0.290$ & $5.09$ & $0.961$
 & \multirow{3}{*}{$7.79$} & \multirow{3}{*}{$2.57$}
 & \multirow{3}{*}{$31.9$} & \multirow{3}{*}{$8.66$} & \multirow{3}{*}{$43.8$} & \multirow{3}{*}{$318$} \\
\cline{2-8}
 & $6.74\times10^{6}$ & $3.72\times10^{-5}$ & $7.71\times10^{-5}$ & $1.80\times10^{-5}$ & $1.12$ & $4.23$ & $3.26$ &&&&&&\\
\cline{2-8}
 & $1.22\times10^{7}$ & $3.16\times10^{-5}$ & $7.28\times10^{-5}$ & $3.44\times10^{-5}$ & $5.71$ & $2.73$ & $4.72$ &&&&&&\\
\hline
\multirow{3}{*}{GeV}
 & $1.67\times10^{9}$ & $2.01\times10^{-6}$ & $1.65\times10^{-7}$ & $5.19\times10^{-7}$ & $1.44$ & $3.97$ & $5.02$
 & \multirow{3}{*}{$7.30$} & \multirow{3}{*}{$2.40$}
 & \multirow{3}{*}{$32.5$} & \multirow{3}{*}{$8.74$} & \multirow{3}{*}{$43.3$} & \multirow{3}{*}{$285$} \\
\cline{2-8}
 & $3.76\times10^{9}$ & $9.30\times10^{-7}$ & $1.18\times10^{-6}$ & $2.31\times10^{-6}$ & $6.07$ & $1.14$ & $1.33$ &&&&&&\\
\cline{2-8}
 & $6.25\times10^{9}$ & $7.18\times10^{-7}$ & $2.25\times10^{-6}$ & $1.69\times10^{-6}$ & $1.21$ & $2.81$ & $0.646$ &&&&&&\\
\hline
\end{tabular}%
}
\end{sidewaystable*}

\bibliographystyle{plain}
\bibliography{references.bib}

@article{Bilenky:1987ty,
    author = "Bilenky, Samoil M. and Petcov, S. T.",
    title = "{Massive Neutrinos and Neutrino Oscillations}",
    doi = "10.1103/RevModPhys.59.671",
    journal = "Rev. Mod. Phys.",
    volume = "59",
    pages = "671",
    year = "1987",
    note = "[Erratum: Rev.Mod.Phys. 61, 169 (1989), Erratum: Rev.Mod.Phys. 60, 575--575 (1988)]"
}

@article{Xing:2011ur,
    author = "Xing, Zhi-zhong",
    title = "{A full parametrization of the 6 X 6 flavor mixing matrix in the presence of three light or heavy sterile neutrinos}",
    eprint = "1110.0083",
    archivePrefix = "arXiv",
    primaryClass = "hep-ph",
    doi = "10.1103/PhysRevD.85.013008",
    journal = "Phys. Rev. D",
    volume = "85",
    pages = "013008",
    year = "2012"
}

@article{Xing:2024gmy,
    author = "Xing, Zhi-zhong and Zhu, Jing-yu",
    title = "{Confronting the seesaw mechanism with neutrino oscillations: A general and explicit analytical bridge}",
    eprint = "2412.17698",
    archivePrefix = "arXiv",
    primaryClass = "hep-ph",
    doi = "10.1016/j.nuclphysb.2025.117041",
    journal = "Nucl. Phys. B",
    volume = "1018",
    pages = "117041",
    year = "2025"
}

@article{Das:2023aic,
    author = "Das, Debashree Priyadarsini and Mishra, Sasmita",
    title = "{Study of neutrinoless double beta decay in the Standard Model extended with sterile neutrinos}",
    eprint = "2310.13353",
    archivePrefix = "arXiv",
    primaryClass = "hep-ph",
    doi = "10.1140/epjc/s10052-024-13024-w",
    journal = "Eur. Phys. J. C",
    volume = "84",
    number = "7",
    pages = "683",
    year = "2024"
}

@article{deGouvea:2022kma,
    author = "de Gouv{\^e}a, Andr{\'e} and Jusino S{\'a}nchez, Giancarlo and Kelly, Kevin J.",
    title = "{Very light sterile neutrinos at NOvA and T2K}",
    eprint = "2204.09130",
    archivePrefix = "arXiv",
    primaryClass = "hep-ph",
    reportNumber = "FERMILAB-PUB-22-247-T, CERN-TH-2022-060",
    doi = "10.1103/PhysRevD.106.055025",
    journal = "Phys. Rev. D",
    volume = "106",
    number = "5",
    pages = "055025",
    year = "2022"
}

@article{Minkowski:1977sc,
    author = "Minkowski, Peter",
    title = "{$\mu \to e\gamma$ at a Rate of One Out of $10^{9}$ Muon Decays?}",
    reportNumber = "Print-77-0182 (BERN)",
    doi = "10.1016/0370-2693(77)90435-X",
    journal = "Phys. Lett. B",
    volume = "67",
    pages = "421--428",
    year = "1977"
}

@article{Gell-Mann:1979vob,
    author = "Gell-Mann, Murray and Ramond, Pierre and Slansky, Richard",
    title = "{Complex Spinors and Unified Theories}",
    eprint = "1306.4669",
    archivePrefix = "arXiv",
    primaryClass = "hep-th",
    reportNumber = "PRINT-80-0576",
    journal = "Conf. Proc. C",
    volume = "790927",
    pages = "315--321",
    year = "1979"
}

@article{KATRIN:2024cdt,
    author = "Aker, Max and others",
    collaboration = "KATRIN",
    title = "{Direct neutrino-mass measurement based on 259 days of KATRIN data}",
    eprint = "2406.13516",
    archivePrefix = "arXiv",
    primaryClass = "nucl-ex",
    doi = "10.1126/science.adq9592",
    journal = "Science",
    volume = "388",
    number = "6743",
    pages = "adq9592",
    year = "2025"
}

@article{KATRIN:2020dpx,
    author = "Aker, M. and others",
    collaboration = "KATRIN",
    title = "{Bound on 3+1 Active-Sterile Neutrino Mixing from the First Four-Week Science Run of KATRIN}",
    eprint = "2011.05087",
    archivePrefix = "arXiv",
    primaryClass = "hep-ex",
    doi = "10.1103/PhysRevLett.126.091803",
    journal = "Phys. Rev. Lett.",
    volume = "126",
    number = "9",
    pages = "091803",
    year = "2021"
}

@article{KATRIN:2022spi,
    author = "Aker, M. and others",
    collaboration = "KATRIN",
    title = "{Search for keV-scale sterile neutrinos with the first KATRIN data}",
    eprint = "2207.06337",
    archivePrefix = "arXiv",
    primaryClass = "nucl-ex",
    doi = "10.1140/epjc/s10052-023-11818-y",
    journal = "Eur. Phys. J. C",
    volume = "83",
    number = "8",
    pages = "763",
    year = "2023"
}

@article{KATRIN:2018oow,
    author = "Mertens, Susanne and others",
    collaboration = "KATRIN",
    title = "{A novel detector system for KATRIN to search for keV-scale sterile neutrinos}",
    eprint = "1810.06711",
    archivePrefix = "arXiv",
    primaryClass = "physics.ins-det",
    doi = "10.1088/1361-6471/ab12fe",
    journal = "J. Phys. G",
    volume = "46",
    number = "6",
    pages = "065203",
    year = "2019"
}

@article{Garnica:2023ccx,
    author = "Garnica, J. C. and Hern{\'a}ndez-Tom{\'e}, G. and Peinado, E.",
    title = "{Charged lepton-flavor violating processes and suppression of nonunitary mixing effects in low-scale seesaw models}",
    eprint = "2302.07379",
    archivePrefix = "arXiv",
    primaryClass = "hep-ph",
    doi = "10.1103/PhysRevD.108.035033",
    journal = "Phys. Rev. D",
    volume = "108",
    number = "3",
    pages = "035033",
    year = "2023"
}

@article{Mohapatra:1979ia,
    author = "Mohapatra, Rabindra N. and Senjanovic, Goran",
    title = "{Neutrino Mass and Spontaneous Parity Nonconservation}",
    reportNumber = "MDDP-TR-80-060, MDDP-PP-80-105, CCNY-HEP-79-10",
    doi = "10.1103/PhysRevLett.44.912",
    journal = "Phys. Rev. Lett.",
    volume = "44",
    pages = "912",
    year = "1980"
}

@article{LSND:2001aii,
    author = "Aguilar, A. and others",
    collaboration = "LSND",
    title = "{Evidence for neutrino oscillations from the observation of $\bar{\nu}_e$ appearance in a $\bar{\nu}_\mu$
 beam}",
    eprint = "hep-ex/0104049",
    archivePrefix = "arXiv",
    doi = "10.1103/PhysRevD.64.112007",
    journal = "Phys. Rev. D",
    volume = "64",
    pages = "112007",
    year = "2001"
}

@article{MiniBooNE:2018esg,
    author = "Aguilar-Arevalo, A. A. and others",
    collaboration = "MiniBooNE",
    title = "{Significant Excess of ElectronLike Events in the MiniBooNE Short-Baseline Neutrino Experiment}",
    eprint = "1805.12028",
    archivePrefix = "arXiv",
    primaryClass = "hep-ex",
    reportNumber = "LA-UR-18-24586, FERMILAB-PUB-18-219-AD-PPD-ND",
    doi = "10.1103/PhysRevLett.121.221801",
    journal = "Phys. Rev. Lett.",
    volume = "121",
    number = "22",
    pages = "221801",
    year = "2018"
}

@article{Planck:2018vyg,
    author = "Aghanim, N. and others",
    collaboration = "Planck",
    title = "{Planck 2018 results. VI. Cosmological parameters}",
    eprint = "1807.06209",
    archivePrefix = "arXiv",
    primaryClass = "astro-ph.CO",
    doi = "10.1051/0004-6361/201833910",
    journal = "Astron. Astrophys.",
    volume = "641",
    pages = "A6",
    year = "2020",
    note = "[Erratum: Astron.Astrophys. 652, C4 (2021)]"
}

@article{NOvA:2018gge,
    author = "Acero, M. A. and others",
    collaboration = "NOvA",
    title = "{New constraints on oscillation parameters from $\nu_e$ appearance and $\nu_\mu$ disappearance in the NOvA experiment}",
    eprint = "1806.00096",
    archivePrefix = "arXiv",
    primaryClass = "hep-ex",
    reportNumber = "FERMILAB-PUB-18-223-ND",
    doi = "10.1103/PhysRevD.98.032012",
    journal = "Phys. Rev. D",
    volume = "98",
    pages = "032012",
    year = "2018"
}

@article{DESI:2016fyo,
    author = "Aghamousa, Amir and others",
    collaboration = "DESI",
    title = "{The DESI Experiment Part I: Science,Targeting, and Survey Design}",
    eprint = "1611.00036",
    archivePrefix = "arXiv",
    primaryClass = "astro-ph.IM",
    reportNumber = "FERMILAB-PUB-16-517-AE",
    month = "10",
    year = "2016"
}

@article{Esteban:2020cvm,
    author = "Esteban, Ivan and Gonzalez-Garcia, M. C. and Maltoni, Michele and Schwetz, Thomas and Zhou, Albert",
    title = "{The fate of hints: updated global analysis of three-flavor neutrino oscillations}",
    eprint = "2007.14792",
    archivePrefix = "arXiv",
    primaryClass = "hep-ph",
    reportNumber = "IFT-UAM/CSIC-112, YITP-SB-2020-21",
    doi = "10.1007/JHEP09(2020)178",
    journal = "JHEP",
    volume = "09",
    pages = "178",
    year = "2020"
}

@article{Wolfenstein:1977ue,
    author = "Wolfenstein, L.",
    title = "{Neutrino Oscillations in Matter}",
    reportNumber = "COO-3066-102",
    doi = "10.1103/PhysRevD.17.2369",
    journal = "Phys. Rev. D",
    volume = "17",
    pages = "2369--2374",
    year = "1978"
}

@article{Mikheyev:1985zog,
    author = "Mikheyev, S. P. and Smirnov, A. Yu.",
    title = "{Resonance Amplification of Oscillations in Matter and Spectroscopy of Solar Neutrinos}",
    journal = "Sov. J. Nucl. Phys.",
    volume = "42",
    pages = "913--917",
    year = "1985"
}

@article{Casas:2001sr,
    author = "Casas, J. A. and Ibarra, A.",
    title = "{Oscillating neutrinos and $\mu \to e, \gamma$}",
    eprint = "hep-ph/0103065",
    archivePrefix = "arXiv",
    reportNumber = "IEM-FT-211-01, OUTP-01-11P, IFT-UAM-CSIC-01-08",
    doi = "10.1016/S0550-3213(01)00475-8",
    journal = "Nucl. Phys. B",
    volume = "618",
    pages = "171--204",
    year = "2001"
}

@article{Ilakovac:1994kj,
    author = "Ilakovac, A. and Pilaftsis, A.",
    title = "{Flavor violating charged lepton decays in seesaw-type models}",
    eprint = "hep-ph/9403398",
    archivePrefix = "arXiv",
    reportNumber = "RAL-94-032, MZ-TH-94-15",
    doi = "10.1016/0550-3213(94)00567-X",
    journal = "Nucl. Phys. B",
    volume = "437",
    pages = "491",
    year = "1995"
}

@article{MEGII:2023ltw,
    author = "Afanaciev, K. and others",
    collaboration = "MEG II",
    title = "{A search for $\mu ^+ \rightarrow \textrm{e}^+ \gamma $ with the first dataset of the MEG~II experiment}",
    eprint = "2310.12614",
    archivePrefix = "arXiv",
    primaryClass = "hep-ex",
    doi = "10.1140/epjc/s10052-024-12416-2",
    journal = "Eur. Phys. J. C",
    volume = "84",
    number = "3",
    pages = "216",
    year = "2024",
    note = "[Erratum: Eur.Phys.J.C 84, 1042 (2024)]"
}

@article{Dolinski:2019nrj,
    author = "Dolinski, Michelle J. and Poon, Alan W. P. and Rodejohann, Werner",
    title = "{Neutrinoless Double-Beta Decay: Status and Prospects}",
    eprint = "1902.04097",
    archivePrefix = "arXiv",
    primaryClass = "nucl-ex",
    doi = "10.1146/annurev-nucl-101918-023407",
    journal = "Ann. Rev. Nucl. Part. Sci.",
    volume = "69",
    pages = "219--251",
    year = "2019"
}

@article{Fernandez-Martinez:2016lgt,
    author = "Fernandez-Martinez, Enrique and Hernandez-Garcia, Josu and Lopez-Pavon, Jacobo",
    title = "{Global constraints on heavy neutrino mixing}",
    eprint = "1605.08774",
    archivePrefix = "arXiv",
    primaryClass = "hep-ph",
    doi = "10.1007/JHEP08(2016)033",
    journal = "JHEP",
    volume = "08",
    pages = "033",
    year = "2016"
}

@article{Abada:2015oba,
    author = "Abada, A. and De Romeri, V. and Teixeira, A. M.",
    title = "{Impact of sterile neutrinos on nuclear-assisted cLFV processes}",
    eprint = "1510.06657",
    archivePrefix = "arXiv",
    primaryClass = "hep-ph",
    reportNumber = "LPT-ORSAY-15-71, PCCF-RI-15-03",
    doi = "10.1007/JHEP02(2016)083",
    journal = "JHEP",
    volume = "02",
    pages = "083",
    year = "2016"
}

@article{Debnath:2023akj,
    author = "Debnath, Hridoy and Fileviez Perez, Pavel",
    title = "{Low scale seesaw mechanism with local lepton number}",
    eprint = "2307.03646",
    archivePrefix = "arXiv",
    primaryClass = "hep-ph",
    doi = "10.1103/PhysRevD.108.075009",
    journal = "Phys. Rev. D",
    volume = "108",
    number = "7",
    pages = "075009",
    year = "2023"
}

@article{Cabrera:2023rcy,
    author = "Cabrera, E. and Cogollo, D. and Pires, C. A. de S.",
    title = "{Naturally low scale type I seesaw mechanism and its viability in the 3-3-1 model with right-handed neutrinos}",
    eprint = "2304.14443",
    archivePrefix = "arXiv",
    primaryClass = "hep-ph",
    doi = "10.1016/j.nuclphysb.2023.116372",
    journal = "Nucl. Phys. B",
    volume = "996",
    pages = "116372",
    year = "2023"
}

@article{CarcamoHernandez:2019kjy,
    author = "C{\'a}rcamo Hern{\'a}ndez, A. E. and Gonz{\'a}lez, Marcela and Neill, Nicol{\'a}s A.",
    title = "{Low scale type I seesaw model for lepton masses and mixings}",
    eprint = "1906.00978",
    archivePrefix = "arXiv",
    primaryClass = "hep-ph",
    doi = "10.1103/PhysRevD.101.035005",
    journal = "Phys. Rev. D",
    volume = "101",
    number = "3",
    pages = "035005",
    year = "2020"
}

@article{Borzumati:2000mc,
    author = "Borzumati, Francesca and Nomura, Yasunori",
    title = "{Low scale seesaw mechanisms for light neutrinos}",
    eprint = "hep-ph/0007018",
    archivePrefix = "arXiv",
    reportNumber = "KEK-TH-701, UT-898",
    doi = "10.1103/PhysRevD.64.053005",
    journal = "Phys. Rev. D",
    volume = "64",
    pages = "053005",
    year = "2001"
}

@article{Dib:2019jod,
    author = "Dib, Claudio and Kovalenko, Sergey and Schmidt, Ivan and Smetana, Adam",
    title = "{Low-scale seesaw from neutrino condensation}",
    eprint = "1904.06280",
    archivePrefix = "arXiv",
    primaryClass = "hep-ph",
    doi = "10.1016/j.nuclphysb.2019.114910",
    journal = "Nucl. Phys. B",
    volume = "952",
    pages = "114910",
    year = "2020"
}

@article{Domcke:2020ety,
    author = "Domcke, Valerie and Drewes, Marco and Hufnagel, Marco and Lucente, Michele",
    title = "{MeV-scale Seesaw and Leptogenesis}",
    eprint = "2009.11678",
    archivePrefix = "arXiv",
    primaryClass = "hep-ph",
    reportNumber = "CERN-TH-2020-158, DESY-20-159, DESY 20-159",
    doi = "10.1007/JHEP01(2021)200",
    journal = "JHEP",
    volume = "01",
    pages = "200",
    year = "2021"
}

@article{SimonsObservatory:2018koc,
    author = "Ade, Peter and others",
    collaboration = "Simons Observatory",
    title = "{The Simons Observatory: Science goals and forecasts}",
    eprint = "1808.07445",
    archivePrefix = "arXiv",
    primaryClass = "astro-ph.CO",
    doi = "10.1088/1475-7516/2019/02/056",
    journal = "JCAP",
    volume = "02",
    pages = "056",
    year = "2019"
}

@article{Euclid:2024imf,
    author = "Archidiacono, M. and others",
    collaboration = "Euclid",
    title = "{Euclid preparation - LIV. Sensitivity to neutrino parameters}",
    eprint = "2405.06047",
    archivePrefix = "arXiv",
    primaryClass = "astro-ph.CO",
    reportNumber = "TTK-24-17",
    doi = "10.1051/0004-6361/202450859",
    journal = "Astron. Astrophys.",
    volume = "693",
    pages = "A58",
    year = "2025"
}

@article{Abazajian:2019eic,
    author = "Abazajian, Kevork and others",
    title = "{CMB-S4 Science Case, Reference Design, and Project Plan}",
    eprint = "1907.04473",
    archivePrefix = "arXiv",
    primaryClass = "astro-ph.IM",
    reportNumber = "FERMILAB-PUB-19-431-AE-SCD",
    month = "7",
    year = "2019"
}

@article{Project8:2022hun,
    author = "Ashtari Esfahani, A. and others",
    collaboration = "Project 8",
    title = "{Tritium Beta Spectrum Measurement and Neutrino Mass Limit from Cyclotron Radiation Emission Spectroscopy}",
    eprint = "2212.05048",
    archivePrefix = "arXiv",
    primaryClass = "nucl-ex",
    doi = "10.1103/PhysRevLett.131.102502",
    journal = "Phys. Rev. Lett.",
    volume = "131",
    number = "10",
    pages = "102502",
    year = "2023"
}

@article{DayaBay:2022orm,
    author = "An, F. P. and others",
    collaboration = "Daya Bay",
    title = "{Precision Measurement of Reactor Antineutrino Oscillation at Kilometer-Scale Baselines by Daya Bay}",
    eprint = "2211.14988",
    archivePrefix = "arXiv",
    primaryClass = "hep-ex",
    doi = "10.1103/PhysRevLett.130.161802",
    journal = "Phys. Rev. Lett.",
    volume = "130",
    number = "16",
    pages = "161802",
    year = "2023"
}

@article{T2K:2021xwb,
    author = "Abe, K. and others",
    collaboration = "T2K",
    title = "{Improved constraints on neutrino mixing from the T2K experiment with $\mathbf{3.13\times10^{21}}$ protons on target}",
    eprint = "2101.03779",
    archivePrefix = "arXiv",
    primaryClass = "hep-ex",
    doi = "10.1103/PhysRevD.103.112008",
    journal = "Phys. Rev. D",
    volume = "103",
    number = "11",
    pages = "112008",
    year = "2021"
}

@article{Esteban:2024eli,
    author = "Esteban, Ivan and Gonzalez-Garcia, M. C. and Maltoni, Michele and Martinez-Soler, Ivan and Pinheiro, Jo{\~a}o Paulo and Schwetz, Thomas",
    title = "{NuFit-6.0: updated global analysis of three-flavor neutrino oscillations}",
    eprint = "2410.05380",
    archivePrefix = "arXiv",
    primaryClass = "hep-ph",
    reportNumber = "IFT-UAM/CSIC-24-140, YITP-SB-2024-24, IPPP/24/64, IPPP/24/64, IFT-UAM/CSIC-24-140, YITP-SB-2024-24",
    doi = "10.1007/JHEP12(2024)216",
    journal = "JHEP",
    volume = "12",
    pages = "216",
    year = "2024"
}

@article{NOvA:2025tmb,
    author = "Abubakar, S. and others",
    collaboration = "NOvA",
    title = "{Precision Measurement of Neutrino Oscillation Parameters with 10 Years of Data from the NOvA Experiment}",
    eprint = "2509.04361",
    archivePrefix = "arXiv",
    primaryClass = "hep-ex",
    reportNumber = "FERMILAB-PUB-25-0619-PPD",
    doi = "10.1103/x53y-2b86",
    journal = "Phys. Rev. Lett.",
    volume = "136",
    number = "1",
    pages = "011802",
    year = "2026"
}

@article{T2K:2025wet,
    author = "Abubakar, S. and others",
    collaboration = "T2K, NOvA",
    title = "{Joint neutrino oscillation analysis from the T2K and NOvA experiments}",
    eprint = "2510.19888",
    archivePrefix = "arXiv",
    primaryClass = "hep-ex",
    reportNumber = "FERMILAB-PUB-25-0132-PPD",
    doi = "10.1038/s41586-025-09599-3",
    journal = "Nature",
    volume = "646",
    number = "8086",
    pages = "818--824",
    year = "2025"
}

@article{JUNO:2015zny,
    author = "An, Fengpeng and others",
    collaboration = "JUNO",
    title = "{Neutrino Physics with JUNO}",
    eprint = "1507.05613",
    archivePrefix = "arXiv",
    primaryClass = "physics.ins-det",
    doi = "10.1088/0954-3899/43/3/030401",
    journal = "J. Phys. G",
    volume = "43",
    number = "3",
    pages = "030401",
    year = "2016"
}

@article{Super-Kamiokande:1998kpq,
    author = "Fukuda, Y. and others",
    collaboration = "Super-Kamiokande",
    title = "{Evidence for oscillation of atmospheric neutrinos}",
    eprint = "hep-ex/9807003",
    archivePrefix = "arXiv",
    reportNumber = "BU-98-17, ICRR-REPORT-422-98-18, UCI-98-8, KEK-PREPRINT-98-95, LSU-HEPA-5-98, UMD-98-003, SBHEP-98-5, TKU-PAP-98-06, TIT-HPE-98-09",
    doi = "10.1103/PhysRevLett.81.1562",
    journal = "Phys. Rev. Lett.",
    volume = "81",
    pages = "1562--1567",
    year = "1998"
}

@article{DUNE:2025sjq,
    author = "Abed Abud, Adam and others",
    collaboration = "DUNE",
    title = "{The DUNE Science Program}",
    eprint = "2503.23291",
    archivePrefix = "arXiv",
    primaryClass = "hep-ex",
    reportNumber = "FERMILAB-PUB-25-0243-LBNF",
    month = "3",
    year = "2025"
}

@article{JUNO:2021vlw,
    author = "Abusleme, Angel and others",
    collaboration = "JUNO",
    title = "{JUNO physics and detector}",
    eprint = "2104.02565",
    archivePrefix = "arXiv",
    primaryClass = "hep-ex",
    doi = "10.1016/j.ppnp.2021.103927",
    journal = "Prog. Part. Nucl. Phys.",
    volume = "123",
    pages = "103927",
    year = "2022"
}

@article{SNO:2002tuh,
    author = "Ahmad, Q. R. and others",
    collaboration = "SNO",
    title = "{Direct evidence for neutrino flavor transformation from neutral current interactions in the Sudbury Neutrino Observatory}",
    eprint = "nucl-ex/0204008",
    archivePrefix = "arXiv",
    doi = "10.1103/PhysRevLett.89.011301",
    journal = "Phys. Rev. Lett.",
    volume = "89",
    pages = "011301",
    year = "2002"
}

@article{NOvA:2007rmc,
    author = "Ayres, D. S. and others",
    collaboration = "NOvA",
    title = "{The NOvA Technical Design Report}",
    reportNumber = "FERMILAB-DESIGN-2007-01",
    doi = "10.2172/935497",
    month = "10",
    year = "2007"
}

@article{Forero:2011pc,
    author = "Forero, D. V. and Morisi, S. and Tortola, M. and Valle, J. W. F.",
    title = "{Lepton flavor violation and non-unitary lepton mixing in low-scale type-I seesaw}",
    eprint = "1107.6009",
    archivePrefix = "arXiv",
    primaryClass = "hep-ph",
    reportNumber = "IFIC-11-38",
    doi = "10.1007/JHEP09(2011)142",
    journal = "JHEP",
    volume = "09",
    pages = "142",
    year = "2011"
}

@article{Flores-Tlalpa:2001vbz,
    author = "Flores-Tlalpa, A. and Hernandez, J. M. and Tavares-Velasco, G. and Toscano, J. J.",
    title = "{Effective Lagrangian description of the lepton flavor violating decays Z ---{\ensuremath{>}} l-+(i) l+-(j)}",
    eprint = "hep-ph/0112065",
    archivePrefix = "arXiv",
    doi = "10.1103/PhysRevD.65.073010",
    journal = "Phys. Rev. D",
    volume = "65",
    pages = "073010",
    year = "2002"
}

@article{Yanagida:1979as,
    author = "Yanagida, Tsutomu",
    editor = "Sawada, Osamu and Sugamoto, Akio",
    title = "{Horizontal gauge symmetry and masses of neutrinos}",
    reportNumber = "KEK-79-18-95",
    journal = "Conf. Proc. C",
    volume = "7902131",
    pages = "95--99",
    year = "1979"
}

@article{Gonzalez-Garcia:2002bkq,
    author = "Gonzalez-Garcia, M. C. and Nir, Yosef",
    title = "{Neutrino Masses and Mixing: Evidence and Implications}",
    eprint = "hep-ph/0202058",
    archivePrefix = "arXiv",
    reportNumber = "CERN-TH-2002-021, WIS-08-02-FEB-DPP",
    doi = "10.1103/RevModPhys.75.345",
    journal = "Rev. Mod. Phys.",
    volume = "75",
    pages = "345--402",
    year = "2003"
}

@article{Maltoni:2003da,
    author = "Maltoni, M. and Schwetz, T. and Tortola, M. A. and Valle, J. W. F.",
    title = "{Status of three neutrino oscillations after the SNO salt data}",
    eprint = "hep-ph/0309130",
    archivePrefix = "arXiv",
    reportNumber = "IFIC-03-42, TUM-HEP-528-03",
    doi = "10.1103/PhysRevD.68.113010",
    journal = "Phys. Rev. D",
    volume = "68",
    pages = "113010",
    year = "2003"
}

@article{Schechter:1980gr,
    author = "Schechter, J. and Valle, J. W. F.",
    title = "{Neutrino Masses in SU(2) x U(1) Theories}",
    reportNumber = "SU-4217-167, COO-3533-167",
    doi = "10.1103/PhysRevD.22.2227",
    journal = "Phys. Rev. D",
    volume = "22",
    pages = "2227",
    year = "1980"
}

@article{MicroBooNE:2021tya,
    author = "Abratenko, P. and others",
    collaboration = "MicroBooNE",
    title = "{Search for an Excess of Electron Neutrino Interactions in MicroBooNE Using Multiple Final-State Topologies}",
    eprint = "2110.14054",
    archivePrefix = "arXiv",
    primaryClass = "hep-ex",
    reportNumber = "FERMILAB-PUB-21-500-ND",
    doi = "10.1103/PhysRevLett.128.241801",
    journal = "Phys. Rev. Lett.",
    volume = "128",
    number = "24",
    pages = "241801",
    year = "2022"
}

@article{Boyarsky:2009ix,
    author = "Boyarsky, Alexey and Ruchayskiy, Oleg and Shaposhnikov, Mikhail",
    title = "{The Role of sterile neutrinos in cosmology and astrophysics}",
    eprint = "0901.0011",
    archivePrefix = "arXiv",
    primaryClass = "hep-ph",
    doi = "10.1146/annurev.nucl.010909.083654",
    journal = "Ann. Rev. Nucl. Part. Sci.",
    volume = "59",
    pages = "191--214",
    year = "2009"
}

@article{Drewes:2013gca,
    author = "Drewes, Marco",
    title = "{The Phenomenology of Right Handed Neutrinos}",
    eprint = "1303.6912",
    archivePrefix = "arXiv",
    primaryClass = "hep-ph",
    reportNumber = "TUM-HEP-881-13",
    doi = "10.1142/S0218301313300191",
    journal = "Int. J. Mod. Phys. E",
    volume = "22",
    pages = "1330019",
    year = "2013"
}

@article{Antusch:2006vwa,
    author = "Antusch, S. and Biggio, C. and Fernandez-Martinez, E. and Gavela, M. B. and Lopez-Pavon, J.",
    title = "{Unitarity of the Leptonic Mixing Matrix}",
    eprint = "hep-ph/0607020",
    archivePrefix = "arXiv",
    reportNumber = "FTUAM-06-8, IFT-UAM-CSIC-06-30",
    doi = "10.1088/1126-6708/2006/10/084",
    journal = "JHEP",
    volume = "10",
    pages = "084",
    year = "2006"
}

@article{Asaka:2005an,
    author = "Asaka, Takehiko and Blanchet, Steve and Shaposhnikov, Mikhail",
    title = "{The nuMSM, dark matter and neutrino masses}",
    eprint = "hep-ph/0503065",
    archivePrefix = "arXiv",
    doi = "10.1016/j.physletb.2005.09.070",
    journal = "Phys. Lett. B",
    volume = "631",
    pages = "151--156",
    year = "2005"
}

@article{Akhmedov:1998qx,
    author = "Akhmedov, Evgeny K. and Rubakov, V. A. and Smirnov, A. Yu.",
    title = "{Baryogenesis via neutrino oscillations}",
    eprint = "hep-ph/9803255",
    archivePrefix = "arXiv",
    reportNumber = "IC-98-22, INR-98-14-T",
    doi = "10.1103/PhysRevLett.81.1359",
    journal = "Phys. Rev. Lett.",
    volume = "81",
    pages = "1359--1362",
    year = "1998"
}

@article{Fukugita:1986hr,
    author = "Fukugita, M. and Yanagida, T.",
    title = "{Baryogenesis Without Grand Unification}",
    reportNumber = "RIFP-641",
    doi = "10.1016/0370-2693(86)91126-3",
    journal = "Phys. Lett. B",
    volume = "174",
    pages = "45--47",
    year = "1986"
}

@article{Bolton:2019pcu,
    author = "Bolton, Patrick D. and Deppisch, Frank F. and Bhupal Dev, P. S.",
    title = "{Neutrinoless double beta decay versus other probes of heavy sterile neutrinos}",
    eprint = "1912.03058",
    archivePrefix = "arXiv",
    primaryClass = "hep-ph",
    doi = "10.1007/JHEP03(2020)170",
    journal = "JHEP",
    volume = "03",
    pages = "170",
    year = "2020"
}

@article{Abdullahi:2022jlv,
    author = "Abdullahi, Asli M. and others",
    title = "{The present and future status of heavy neutral leptons}",
    eprint = "2203.08039",
    archivePrefix = "arXiv",
    primaryClass = "hep-ph",
    reportNumber = "FERMILAB-CONF-22-184-T-V",
    doi = "10.1088/1361-6471/ac98f9",
    journal = "J. Phys. G",
    volume = "50",
    number = "2",
    pages = "020501",
    year = "2023"
}

@article{Blennow:2023mqx,
    author = "Blennow, Mattias and Fern{\'a}ndez-Mart{\'\i}nez, Enrique and Hern{\'a}ndez-Garc{\'\i}a, Josu and L{\'o}pez-Pav{\'o}n, Jacobo and Marcano, Xabier and Naredo-Tuero, Daniel",
    title = "{Bounds on lepton non-unitarity and heavy neutrino mixing}",
    eprint = "2306.01040",
    archivePrefix = "arXiv",
    primaryClass = "hep-ph",
    reportNumber = "IFT-UAM/CSIC-23-60, FTUV-23-0531.7594, IFIC/23-19",
    doi = "10.1007/JHEP08(2023)030",
    journal = "JHEP",
    volume = "08",
    pages = "030",
    year = "2023"
}

@article{Escrihuela:2015wra,
    author = "Escrihuela, F. J. and Forero, D. V. and Miranda, O. G. and Tortola, M. and Valle, J. W. F.",
    title = "{On the description of nonunitary neutrino mixing}",
    eprint = "1503.08879",
    archivePrefix = "arXiv",
    primaryClass = "hep-ph",
    reportNumber = "IFIC-15-14",
    doi = "10.1103/PhysRevD.92.053009",
    journal = "Phys. Rev. D",
    volume = "92",
    number = "5",
    pages = "053009",
    year = "2015",
    note = "[Erratum: Phys.Rev.D 93, 119905 (2016)]"
}

@article{Klop:2014ima,
    author = "Klop, N. and Palazzo, A.",
    title = "{Imprints of CP violation induced by sterile neutrinos in T2K data}",
    eprint = "1412.7524",
    archivePrefix = "arXiv",
    primaryClass = "hep-ph",
    doi = "10.1103/PhysRevD.91.073017",
    journal = "Phys. Rev. D",
    volume = "91",
    number = "7",
    pages = "073017",
    year = "2015"
}

@article{Parke:2015goa,
    author = "Parke, Stephen and Ross-Lonergan, Mark",
    title = "{Unitarity and the three flavor neutrino mixing matrix}",
    eprint = "1508.05095",
    archivePrefix = "arXiv",
    primaryClass = "hep-ph",
    reportNumber = "IPPP-15-53, DCPT-15-106, FERMILAB-PUB-15-363-T",
    doi = "10.1103/PhysRevD.93.113009",
    journal = "Phys. Rev. D",
    volume = "93",
    number = "11",
    pages = "113009",
    year = "2016"
}

@article{Blennow:2016jkn,
    author = "Blennow, Mattias and Coloma, Pilar and Fernandez-Martinez, Enrique and Hernandez-Garcia, Josu and Lopez-Pavon, Jacobo",
    title = "{Non-Unitarity, sterile neutrinos, and Non-Standard neutrino Interactions}",
    eprint = "1609.08637",
    archivePrefix = "arXiv",
    primaryClass = "hep-ph",
    reportNumber = "IFT-UAM-CSIC-16-090, FTUAM-16-35, FERMILAB-PUB-16-400-T",
    doi = "10.1007/JHEP04(2017)153",
    journal = "JHEP",
    volume = "04",
    pages = "153",
    year = "2017"
}

\end{document}